\documentclass[letterpaper, 10 pt, conference]{ieeeconf}
\usepackage{url}
\usepackage{epsfig}
\usepackage{amsmath}
\usepackage{amssymb}
\usepackage{colortbl}
\usepackage{booktabs}
\usepackage{etoolbox}
\usepackage{rotating}

\newcommand{\nsj}[1]{%
 \ifstrequal{#1}{SJ0084}{1}{}%
 \ifstrequal{#1}{SJ0087}{2}{}%
 \ifstrequal{#1}{SJ0088}{3}{}%
 \ifstrequal{#1}{SJ0085}{4}{}%
 \ifstrequal{#1}{SJ0096}{5}{}%
 \ifstrequal{#1}{SJ0082}{6}{}%
} 

\newcommand{\sj}[1]{%
 \ifstrequal{#1}{SJ0084}{S1}{}%
 \ifstrequal{#1}{SJ0087}{S2}{}%
 \ifstrequal{#1}{SJ0088}{S3}{}%
 \ifstrequal{#1}{SJ0085}{S4}{}%
 \ifstrequal{#1}{SJ0096}{S5}{}%
 \ifstrequal{#1}{SJ0082}{S6}{}%
} 

\newcommand{\fsj}[1]{%
 \ifstrequal{#1}{SJ0084}{Subject 1}{}%
 \ifstrequal{#1}{SJ0087}{Subject 2}{}%
 \ifstrequal{#1}{SJ0088}{Subject 3}{}%
 \ifstrequal{#1}{SJ0085}{Subject 4}{}%
 \ifstrequal{#1}{SJ0096}{Subject 5}{}%
 \ifstrequal{#1}{SJ0082}{Subject 6}{}%
}

\IEEEoverridecommandlockouts                              
                                                          
\definecolor{grey9}{rgb}{0.9,0.9,0.9}

\title{\LARGE \bf
Electrocorticogram encoding of upper extremity movement trajectories}

\author{Po T. Wang$^{3}$, Christine E. King$^{3}$, Andrew Schombs$^{3}$, Jack J. Lin$^{2}$, Mona Sazgar$^{2}$, Frank P.K. Hsu$^{4}$,\\
 Susan J. Shaw$^{5,6}$, David E. Millett$^{5,6}$, Charles Y. Liu$^{7,8}$, Luis A. Chui$^{2}$, Zoran Nenadic$^{3,9}$ and An H. Do$^{1,2}$
\thanks{Work supported by the National Science Foundation (Award \#1134575)}
\thanks{$^{3}$Department of Biomedical Engineering, UCI, CA, USA
        {\tt\small ptwang@uci.edu, znenadic@uci.edu}}%
 \thanks{$^{4}$Department of Neurosurgery, UCI, Irvine, CA, USA}%
 \thanks{$^{5}$Department of Neurology, Rancho Los Amigos National Rehabilitation Center (RLANRC), Downey, CA, USA}%
 \thanks{$^{6}$Department of Neurology, University of Southern California (USC), Los Angeles, CA, USA}%
 \thanks{$^{7}$Department of Neurosurgery, RLANRC, Downey, CA, USA}%
 \thanks{$^{8}$Department of Neurosurgery, USC, Los Angeles, CA, USA}%
 \thanks{$^{9}$Department of Electrical Engineering and Computer Science, UCI, Irvine, CA, USA}%
\thanks{$^{2}$Department of Neurology, University of California, Irvine (UCI), Irvine, CA, USA
        {\tt\small and@uci.edu}}%
}

\begin{document}

\maketitle
\thispagestyle{empty}
\pagestyle{empty}

\begin{abstract}

Electrocorticogram (ECoG)-based brain computer interfaces (BCI) can potentially control upper extremity prostheses to restore independent function to paralyzed individuals. However, current research is mostly restricted to the offline decoding of finger or 2D arm movement trajectories, and these results are modest. This study seeks to improve the fundamental understanding of the ECoG signal features underlying upper extremity movements to guide better BCI design. Subjects undergoing ECoG electrode implantation performed a series of elementary upper extremity movements in an intermittent flexion and extension manner. It was found that movement velocity, $\dot\theta$, had a high positive (negative) correlation with the instantaneous power of the ECoG high-$\gamma$ band (80-160 Hz) during flexion (extension). Also, the correlation was low during idling epochs. Visual inspection of the ECoG high-$\gamma$ band revealed power bursts during flexion/extension events that have a waveform that strongly resembles the corresponding flexion/extension event as seen on $\dot\theta$. These high-$\gamma$ bursts were present in all elementary movements, and were spatially distributed in a somatotopic fashion. Thus, it can be concluded that the high-$\gamma$ power of ECoG strongly encodes for movement trajectories, and can be used as an input feature in future BCIs.

\end{abstract}

\section{Introduction}
Brain-computer interface (BCI)-controlled upper extremity prostheses are a much sought-after application to restore upper extremity function and independence after paralyzing conditions such as cervical spinal cord injury, subcortical stroke, or brainstem lesions. 
Recently, there has been a growing interest in using electrocorticogram (ECoG) as a long-term signal acquisition platform for BCI-control of upper extremity prostheses. Several studies have shown that ECoG signals can be used to decode movement trajectories of the arm and fingers, thereby indicating that the ECoG-based BCI platform for upper extremity prosthesis control is promising. Studies such as \cite{kjmiller:09, jkubanek:09, nliang:12, zwang:11, sacharya:10, hlbenz:12} used local motor potentials (LMPs) and/or the high-$\gamma$ band of ECoG signals to decode trajectories of repetitive finger or arm movements \cite{gschalk:07,tpistohl:08,jcsanchez:08}. The maximum correlation coefficients between the actual and decoded finger trajectories averaged across all subjects within each study ranged between 0.32 -- 0.64.  
Similarly, the correlation between the actual and decoded 2D arm trajectory was 0.3  in~\cite{tpistohl:08}, and varied from 0.50 -- 0.62 in~\cite{jcsanchez:08}. 

The development of an ECoG-based BCI-controlled upper extremity prosthesis to restore motor function and independence to paralyzed individuals must still overcome many limitations. First, with the exception of \cite{kjmiller:09} and \cite{zwang:11}, the existing decoders were unable to accurately predict idling periods, or these idling periods were completely omitted. Hence, it remains unclear how well idling periods can be decoded from ECoG signals for multiple degrees of freedom (DOF). Second, the ability to decode movement trajectories has mostly been studied in the context of repetitive movements. In everyday life, however, intermittent movements of upper extremities are much more common, so it remains unclear if existing decoders can be generalized to these types of movements. Third, the majority of ECoG decoding studies have focused on finger \cite{jkubanek:09, nliang:12, zwang:11, sacharya:10, hlbenz:12} or 2D arm movement trajectories \cite{gschalk:07,tpistohl:08, jcsanchez:08}. However, since activities of daily living require many unique configurations of upper extremities, a ECoG-based BCI must be able to decode at least 6 DOF in order to control an upper extremity prosthesis and restore independence to a user \cite{dpromilly:94}.
Furthermore, the BCI must be able to control each DOF with high accuracy. Unfortunately, the moderate decoding accuracies reported in the current literature may not be sufficient for online BCI control of an upper extremity prosthesis.

To address the above limitations and unknowns, a better fundamental understanding of how ECoG encodes upper extremity movements is required. This may reveal more salient features underlying upper extremity movements, and may lead to the design of superior decoding algorithms.
In order to build a viable ECoG-based BCI system for upper extremity prosthesis control,
there must exist robust ECoG signal features which encode information that answers the following questions:  1. which joint(s) are in the idling or movement state?; and 2. for joints that are moving, what are their trajectories? In this exploratory study, the authors examine the time-frequency characteristics of ECoG signals during 6 elementary upper extremity movements to increase the fundamental understanding of ECoG motor encoding.

\section{Methods}\label{sec:methods}

\subsection{Overview}
In order to examine how ECoG signals encode for movement trajectories at individual joints in the upper extremities, subjects undergoing subdural electrode implantation over primary motor cortex were asked to perform a series of elementary upper extremity movements while their ECoG signals were recorded. Each elementary movement was performed in a continuous oscillation of flexion and extension, and again with intermittent flexion and extension (intervened by a rest period in between each flexion and extension). The resulting signal was then visually and quantitatively compared to kinematic parameters (e.g. position and velocity). Various analysis techniques were applied in order to characterize the relationship between ECoG signals and trajectory.

\subsection{Subjects and ECoG Signal Acquisition}\label{sec:saatdc}
The study was approved by the Institutional Review Boards of the University of California, Irvine and the Rancho Los Amigos National Rehabilitation Center. Subjects were recruited from a patient population undergoing temporary subdural electrode implantation for epilepsy surgery evaluation. Subject selection was limited to those with electrodes placed over the upper extremity representation area of the primary motor cortex (M1). Up to 64 channels of ECoG data were recorded using a pair of linked Nexus-32 bioamplifiers (Mind Media, Roermond-Herten, The Netherlands), and signals were acquired at 2048 Hz sample rate with common average referencing.

\subsection{Task}\label{sec:tasks}
The subjects performed six elementary arm movements on the side contralateral to their ECoG electrode implant \cite{ahdo:13}: \textbf{1.}~pincer grasp and release (PG); \textbf{2.} wrist flexion and extension (W); \textbf{3.} forearm pronation and supination (PS), \textbf{4.} elbow flexion and extension (E); \textbf{5.} shoulder forward flexion and extension (SFE); \textbf{6.} shoulder internal and external rotation (SR). The trajectories of PG and W were measured by a custom-made electrogoniometer~\cite{ptwang:11}, while the movement trajectories of PS, E, SFE, and SR were measured by a gyroscope (Wii Motion Plus, Nintendo, Kyoto, Japan). The trajectory signals, including position, $\theta(t)$, and velocity, $\dot\theta(t)$, were acquired using an integrated microcontroller unit (Arduino, Smart Projects, Turin, Italy). ECoG data were synchronized with the trajectory signals using a common pulse train sent to both acquisition systems.

The above elementary movements were performed sequentially from \textbf{1} to \textbf{6}. Prior to each movement, the appropriate physical sensor was mounted and calibrated using conventional goniometry at $10^{\circ}$ intervals throughout the joint's range of motion. Two types of motor paradigms were performed--an intermittent and a continuous paradigms. Note that except for SFE movement type, subjects' arms were positioned in such a way that these movements were performed in a gravity neutral manner.

\subsubsection{Intermittent Motor Paradigm}
Subjects performed intermittent, alternating flexion and extension movements. A flexion movement was performed until the end of the range of motion and was followed by a idling period (while in the fully flexed position) for 3-5 seconds. Subjects then extended to the end of the range of motion and idled in this fully extended position for 3-5 seconds. This was repeated for a total of 25 cycles for each elementary movement. 

\subsubsection{Continuous Motor Paradigm} 
Each elementary movement type was performed in a continuous oscillatory manner, in which flexion is followed immediately by extension with no resting period in between. Four sets of 25 flexion/extension continuous cycles were performed, with each set intervened by a 20-30-second long resting period.

\subsection{Time-Frequency Analysis}\label{sec:analysis}
Modulation of the low frequency bands in EEG and ECoG is believed to represent the activity of thalamocortical tracts to the sensorimotor cortex as opposed to the activity within the motor cortex (M1). On the other hand, the high-$\gamma$ band in ECoG likely represents the activity of local cortico-cortico tracts of the motor related cortices (e.g. M1, supplementary motor area, premotor area), and therefore may contain information on movement trajectories. Spectral analysis in prior studies \cite{kjmiller:07, kjmiller:09} described that high $\gamma$ power increases during movement when compared to rest. However, the exact nature of movement trajectory encoding within the ECoG high-$\gamma$ band is not yet clear. Here, the authors explore the high-$\gamma$ band encoding by examining the temporal relationship between the $\gamma$-band power and trajectory. This is performed by first calculating the ECoG instantaneous power in the high-$\gamma$ band:
\begin{equation}\label{eq:instantaneouspower}
P_{n}(t) = f(x^{2}_{n}(t))
\end{equation}
where $x_{n}(t)$ is the bandpass filtered ECoG signal in the high-$\gamma$ band at channel $n$ and $P_{n}(t)$ is its power, enveloped by a low-pass filter, $f(\cdot)$. In order to determine the appropriate frequency range for the high-$\gamma$ frequency band, spectrograms were first plotted for each channel across all subjects to help visualize the presence of high-$\gamma$ modulation during epochs of movement and idling. Inspection of these spectrograms were used to empirically determine the parameters of the bandpass filter in Eqn.~\ref{eq:instantaneouspower}.

Subsequently, $P_n(t)$ during the intermittent motor tasks were segmented into flexion, extension, and idle epochs based on $\dot{\theta}(t)$. The cross-correlations between $P_n(t)$ and $\dot{\theta}(t)$ were then calculated during flexion, extension, and idling epochs. The cross-correlations during flexion and extension epochs were lag-optimized independently, while idling cross-correlations were calculated at zero lag. The procedure was repeated for all channels and for all 6 elementary movements in all subjects. 

Further exploration was then performed to determine if any difference exists in $P(t)$ between flexion and extension underlying each movement type. To this end, spectral energies of the high-$\gamma$ band will be calculated during the intermittent motor paradigm. Since the flexion and extension movements were separated by a sufficiently long idling period (3-5 s), ECoG signals encoding these directions were expected to be well separated, and thus their energies should be well defined. For each movement type, the $P(t)$ signals were first standardized to median absolute deviations (MAD) of the idle epochs, such that all idle epochs for any electrode, movement type and paradigms and across subjects have MAD=1. 
Next, the area above 3$\times$MAD and below the $P(t)$ signal corresponding to flexion and extension were calculated and averaged across all epochs. Finally, the spectral energies between flexion and extension epochs were compared using an unpooled t-test.

Similarly, the lag-optimized cross-correlation between $P(t)$ and $\dot{\theta}(t)$ was calculated for all movement types for all subjects for the continuous motor paradigm. Note that for the continuous movement paradigm, only ``moving'' (flexions and extensions) and ``idling'' were separated. Also, the lags were only optimized using the moving epochs, whereas the correlations during the idling epochs were calculated at the same lag as those of the moving epochs.

Due to the periodicity of the continuous movement paradigm, cross-correlation analysis was susceptible to aligning $P(t)$ to $\dot{\theta}$ over more than one cycle away.
A closer examination was necessary to explore the time delays between $P(t)$ and $\dot{\theta}$ for the continuous movement paradigm.
To this end, $P(t)$ signals from the top correlating channels in the M1 area were plotted synchronously with $\dot{\theta}$ and were visually inspected for any corresponding features. The delays between the $P(t)$ features and corresponding $\dot{\theta}$ peaks were measured. The delays were analyzed statistically to reveal their underlying temporal relationship. The averaged $P(t)$ centered around $\dot{\theta}$ maxima and minima were also calculated to verify the results of the visual inspection.

To explore similarities and differences between the continuous and intermittent paradigms, the same is also performed for intermittent paradigm.

\section{Results}

Six subjects undergoing subdural electrode implantation for epilepsy surgery evaluation were recruited for this study. Their demographic data are summarized in Table \ref{tab:demographics}. Most subjects were able to perform both the intermittent and continuous movement paradigms for all 6 elementary movement types. Due to time constraints, \fsj{SJ0085} was only able to perform PG movements, and \fsj{SJ0082} was only able to perform the continuous paradigm for PG, W, PS, and E. Their ECoG data were then analyzed as described in Section \ref{sec:methods}.

\begin{table*}[!htpb]
\caption{Subject Demographics}
\label{tab:demographics}
\centering
\begin{tabular}{cccl}
\toprule
Subject & Age & Gender & Electrode Location \\ 
\midrule
\nsj{SJ0084} & 27 & F & 6$\times$8 right frontal-parietal grid \\ 
\nsj{SJ0087} & 49 & F & 8$\times$8 left frontal-temporal grid, 1$\times$6 posterior frontal-anterior parietal strip \\ 
\nsj{SJ0088} & 22 & M & 4$\times$5 left frontal-parietal grid, two 2$\times$5 frontal-temporal strips \\ 
\nsj{SJ0085} & 35 & F & 1$\times$6 right frontal-parietal strip \\
\nsj{SJ0096} & 23 & F & 8$\times$8 right frontal-temporal-parietal grid \\
\nsj{SJ0082} & 20 & F & 8$\times$8 left frontal-temporal grid, 1$\times$6 frontal-parietal strip \\
\bottomrule
\end{tabular} 
\end{table*}

\subsection{Spectrogram}
\begin{figure*}
\label{spectrogram}
\centering
\includegraphics[width=0.9\linewidth]{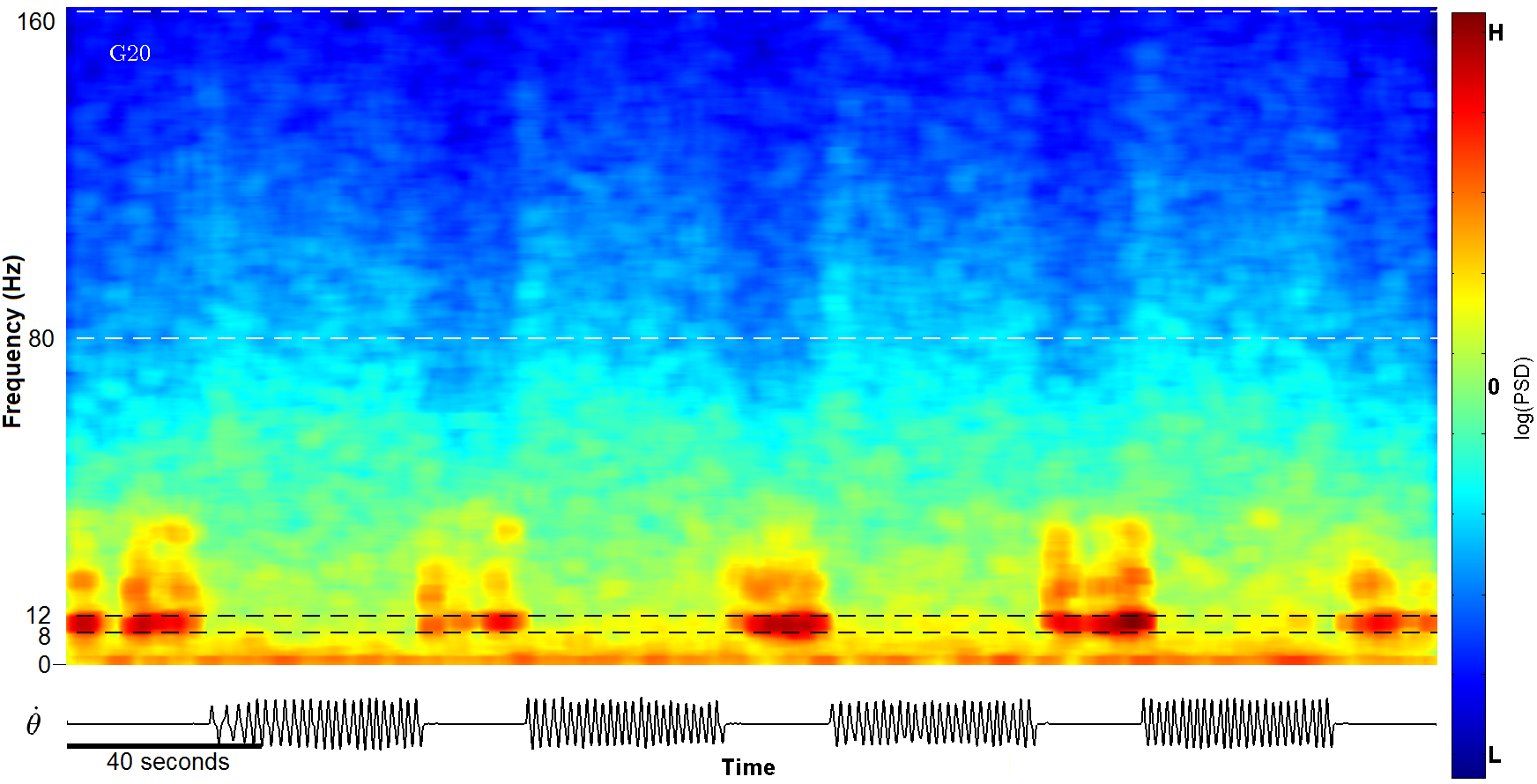}
\caption{
Spectrogram during the PG continuous movement task for \fsj{SJ0084}. 
$\dot{\theta}$ shows idling and movement epochs, where a positive (negative) deflection represents flexion (extension), respectively. 
The spectrogram, with a 4 Hz $\times$ 5 s Mexican-hat window, shows the log power spectral density (PSD) from 0 to 160 Hz for the duration of the movement task.
Note the increased power in the high-$\gamma$ band during the movement epochs. On the other hand, the $\mu$ and $\beta$ bands show increased power during idling epochs.
}
\label{fig:spectrogram}
\end{figure*}

In all subjects, visual inspection of the spectrograms revealed an obvious presence of increased high-$\gamma$ power during movement epochs and decreased power during idling epochs. It was observed  that high-$\gamma$ power in the range of 60-160 Hz was the most common feature during movement epochs across all subjects. Since there was considerable noise at 60 Hz, the frequency range for $P(t)$ in Eqn.~\ref{eq:instantaneouspower} was set to 80-160 Hz. Note that low-frequency bands were observed to modulate in an opposite manner. A representative spectrogram can be seen in Fig. \ref{fig:spectrogram}.
 
\subsection{Time-Frequency Analysis}

Visual inspection of the $P(t)$ signals revealed a burst of power that was time-locked to every flexion or extension event during the intermittent paradigm. The $P(t)$ signal during idling appeared noisy and chaotic. Additionally, the waveform of $P(t)$ during these bursts closely matched the visual appearance of the flexion and extension waveforms seen in $\dot\theta(t)$. A representative set of tracings can be seen in Fig. \ref{fig:intermittentwaveform}. 

\subsubsection{Intermittent Motor Paradigm}
To quantify the similarity between $P(t)$ and $\dot\theta(t)$ during the intermittent paradigm, the cross-correlation between $P(t)$ and $\dot\theta(t)$ during flexion, extension, and idling epochs were calculated (see Section \ref{sec:analysis}). Based on the visual appearance of $P(t)$ and $\dot\theta(t)$, the results were as expected: high positive cross-correlation for flexion epochs, low correlation for idle epochs, and high negative cross-correlation for extension epochs.
The electrodes located over M1 were ranked based on the correlation pattern, and the top 1-3 electrodes were reported in Table \ref{tab:intermittent}. Note that electrodes straddling the central sulcus were also considered in this ranking report.
The optimal latencies were generally small and did not consistently lead or lag the onset of movements. 
However, results from the optimal lag search indicate that correlations generally start strengthening before the onset of movement as demonstrated in the correlation-lag diagrams in Fig. \ref{fig:subject1lagcorrelation}.
To better characterize the temporal relationship between $P(t)$ and $\dot{\theta}$, the timestamps of each peak of the power burst and of the corresponding velocity extrema were manually identified. The time difference of each pair of timestamps were calculated, and the results are summarized in Fig.~\ref{fig:phaseshift2}. 
There were no consistent lead or lag times for the small joints.
On the other hand, $P(t)$ for large joints tend to lead $\dot{\theta}$, such as the SFE flexion movements that showed a 500 ms lead time across all subjects.
When averaged, $P(t)$ peaks also lead for large joints, and the widths of the peaks are also wider than those of the small joints (see Fig.~\ref{afig:amp}).

\begin{figure*}[!ht]
\centering
\includegraphics[width=1.0\linewidth]{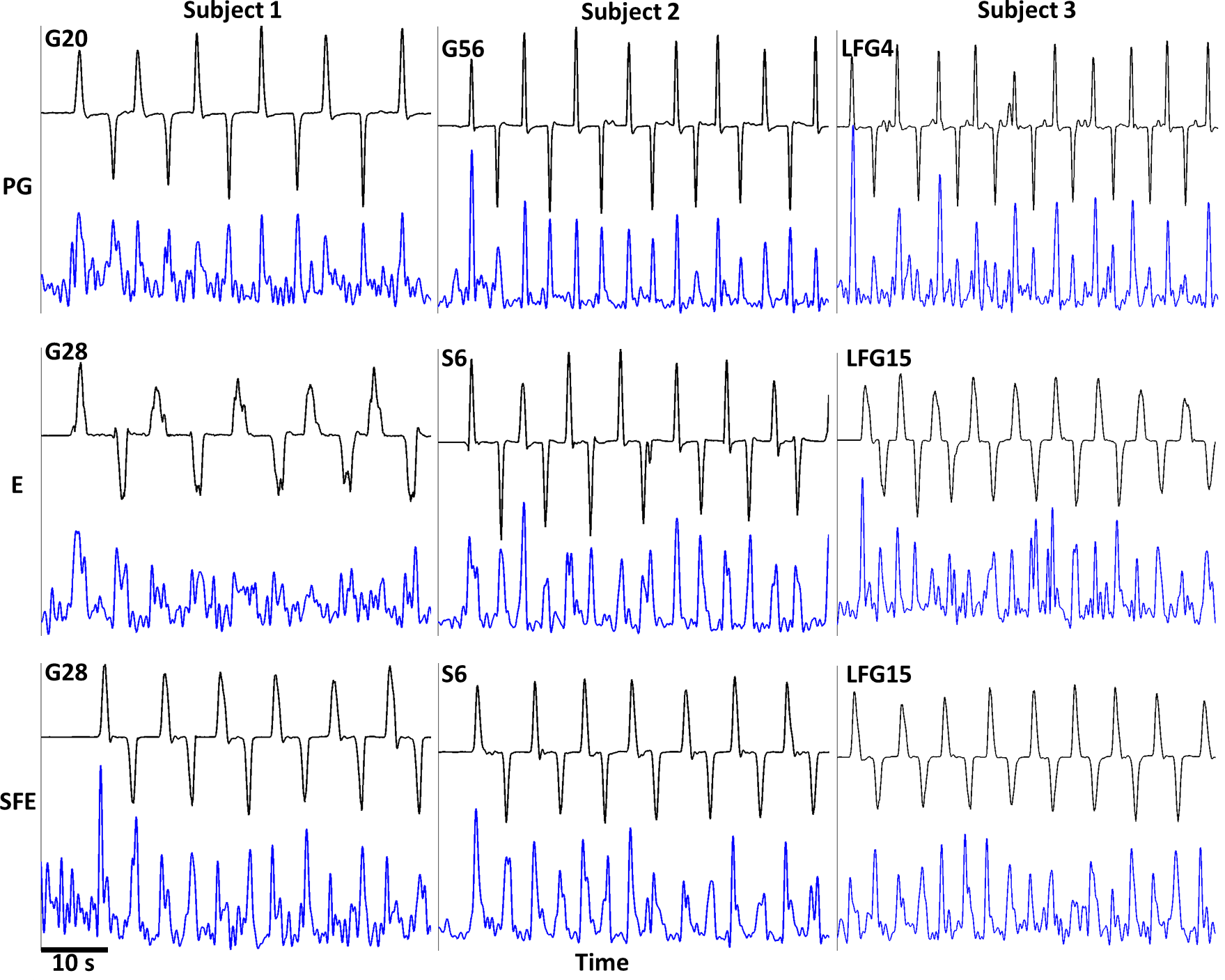}
\caption{
A representative segment of $P(t)$ (blue trace) and corresponding $\dot\theta(t)$ (black) at the best M1 electrode for each Subject. These are shown for elementary movement types PG, E, SFE.
}
\label{fig:intermittentwaveform}
\end{figure*}

The spectral energy of high-$\gamma$ power bursts during flexion and extension were quantified, and the results are summarized in Table \ref{tab:intermittent}. A disparity between the flexion and extension energies were observed in the relevant M1 electrodes across all available subjects in all movement types. In particular, PG movements were typically represented by at least one electrode with a very strong difference between flexion and extension energies across all available subjects (e.g. \fsj{SJ0088}, PG movement in Fig. \ref{fig:intermittentwaveform}). 
However, no systematic preference of direction was seen in any movement types or electrodes.

\subsubsection{Continuous Motor Paradigm} 
\begin{figure*}[]
\centering
\includegraphics[width=\linewidth]{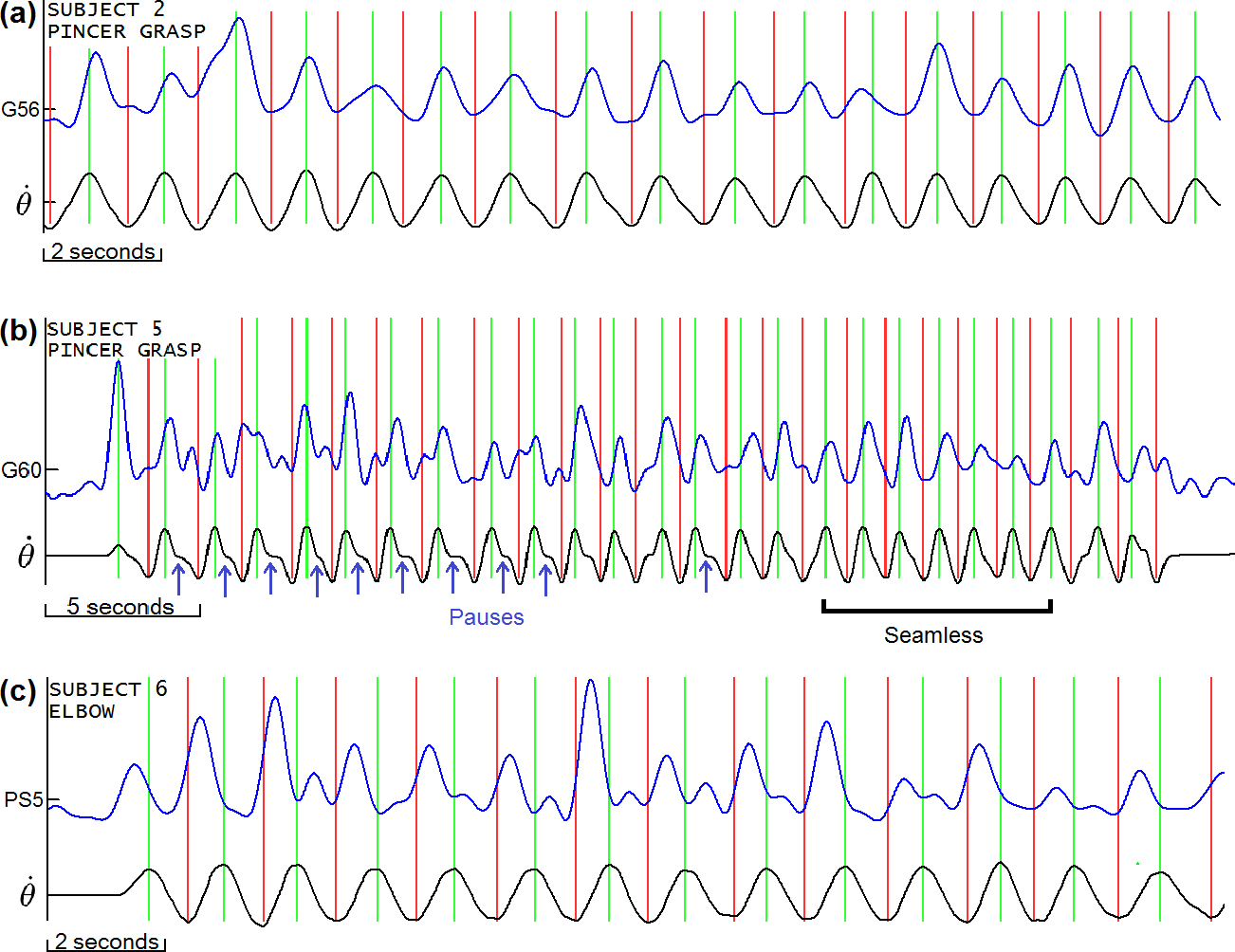}
\caption{
(a) $P(t)$ and $\dot{\theta}$ plots for \fsj{SJ0087} during pincer grasp continuous movement. Vertical lines denote the times of maximum flexion (green) and extension (red) velocities ($\dot{\theta}$ peaks) and assisted in finding the corresponding $P(t)$ peak. Triangles denote manually marked flexion peaks sufficiently above the baseline amplitude. Corresponding high-$\gamma$ power bursts can be seen on $P(t)$, slightly lagging behind flexion peaks. Also note the weaker, almost non-existent, extension peaks (not marked).
(b) $P(t)$ and $\dot{\theta}$ plots for \fsj{SJ0096} during elbow continuous movement. Similarly, corresponding high-$\gamma$ power bursts can be seen on $P(t)$. For the first 2/3 of the movement epoch, the subject paused briefly after each flexion. The extension peaks were thus visible. On the other hand, during "Non-Paused" region where the subject did not pause, the extension peaks disappeared.
(c) $P(t)$ and $\dot{\theta}$ plots for \fsj{SJ0082} during elbow continuous movement. Corresponding high-$\gamma$ power bursts can be seen on $P(t)$, but instead lead each flexion peak by 1/4-cycle.
}
\label{fig:bigsmalljoint}
\end{figure*}

During continuous movements, multiple patterns of high-$\gamma$ bursts in the $P(t)$ signals were observed. 
First, idling was characterized by low powered, noisy and chaotic $P(t)$ signal, which is similar to idling during the intermittent motor tasks.
Next, the flexion movement events coincided with the high-$\gamma$ power bursts, similar to the pattern observed in the intermittent movements above. 
However, not all extension events were coupled with power bursts. 
More specifically, extension-associated power bursts are mostly absent or very weak in small joint movements such as PG, W, and PS (see Figs.~\ref{fig:bigsmalljoint}a and \ref{afig:amp}). The numbers of visually detectable extension-associated power bursts are presented in Table~\ref{tab:numidentified}. An exception to this is observed when subjects unintentionally fail to maintain continuous movement and introduce a brief pause between each flexion and extension movement, such as depicted in Fig.~\ref{fig:bigsmalljoint}b. In this case, the extension peaks became more noticeable.
On the other hand, for large joint movements such as E, SFE, and SR, both flexion and extension peaks are visible even while the subject engaged in seamless movements (Figs.~\ref{fig:bigsmalljoint}c and \ref{afig:amp}).
Finally, a consistent phase shift was also observed between the power bursts and the movement events in large joint movements (compare Figs.~\ref{fig:bigsmalljoint}a and c).

To quantify these observed patterns relating $P(t)$ to $\dot{\theta}$, a series of analyses were performed as described in Methods. 
First, a lag-optimized cross-correlation analysis was performed. Electrodes located over M1 were ranked based on the correlation, and the top 1-3 electrodes were reported in Table~\ref{tab:continuous}. However, the correlations were much weaker (average 0.38) compared to those during intermittent movements (average 0.56).

\begin{figure*}[!ht]
\centering
\includegraphics[width=1.0\linewidth]{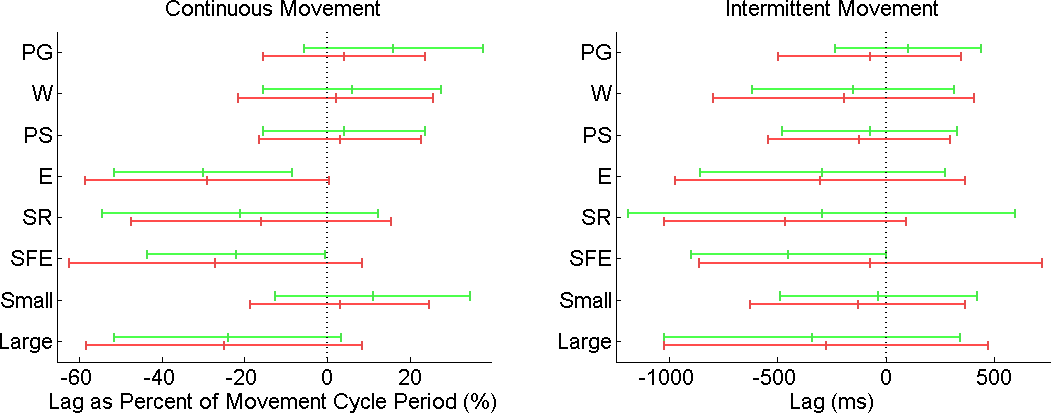}
\caption{
Phase-shift analysis of small and large joint movements for both continuous and intermittent paradigms. 
Lags denote the time between the $P(t)$ peaks and the $\dot{\theta}$ peaks and are represented as percentages of one movement cycle duration for the continuous paradigm and as time for the intermittent paradigm ($P(t)$ leads $\dot{\theta}$ when lag is negative). 
The lags for each movement are divided between flexion/supination/internal rotation (green bars) and extension/pronation/external rotation (red bars).
The vertical bars in each lag denote means and $\pm$1.96$\times$standard deviations.
Not all sessions have clearly discernible $P(t)$ peaks and the number of identified peaks are summarized in Table~\ref{tab:numidentified}.
}
\label{fig:phaseshift2}
\end{figure*}

Second, the temporal relation between $P(t)$ and $\dot{\theta}$ was determined. 
Where visually identifiable, the high-$\gamma$ power bursts were marked (see Fig.~\ref{fig:bigsmalljoint}), and the lag times relative to the corresponding flexion/extension events (based on $\dot{\theta}$) were measured. The results are summarized in Fig.~\ref{fig:phaseshift2}. 
It was found that the time at which the $P(t)$ peak occurred are different for small and large joints. For small joints, the $P(t)$ peaks occurred 11$\pm$12 \% of the movement periods after $\dot{\theta}$ peaks. On the other hand, $P(t)$ peaks of large joints lead the $\dot{\theta}$ peaks by 24$\pm$14 \% of the periods, with the lead times in the E and SFE movements being significant. The individual details are provided in Appendix Table~\ref{atab:phaseshift2}.

\begin{figure*}[!ht]
\centering
\includegraphics[width=0.9\linewidth]{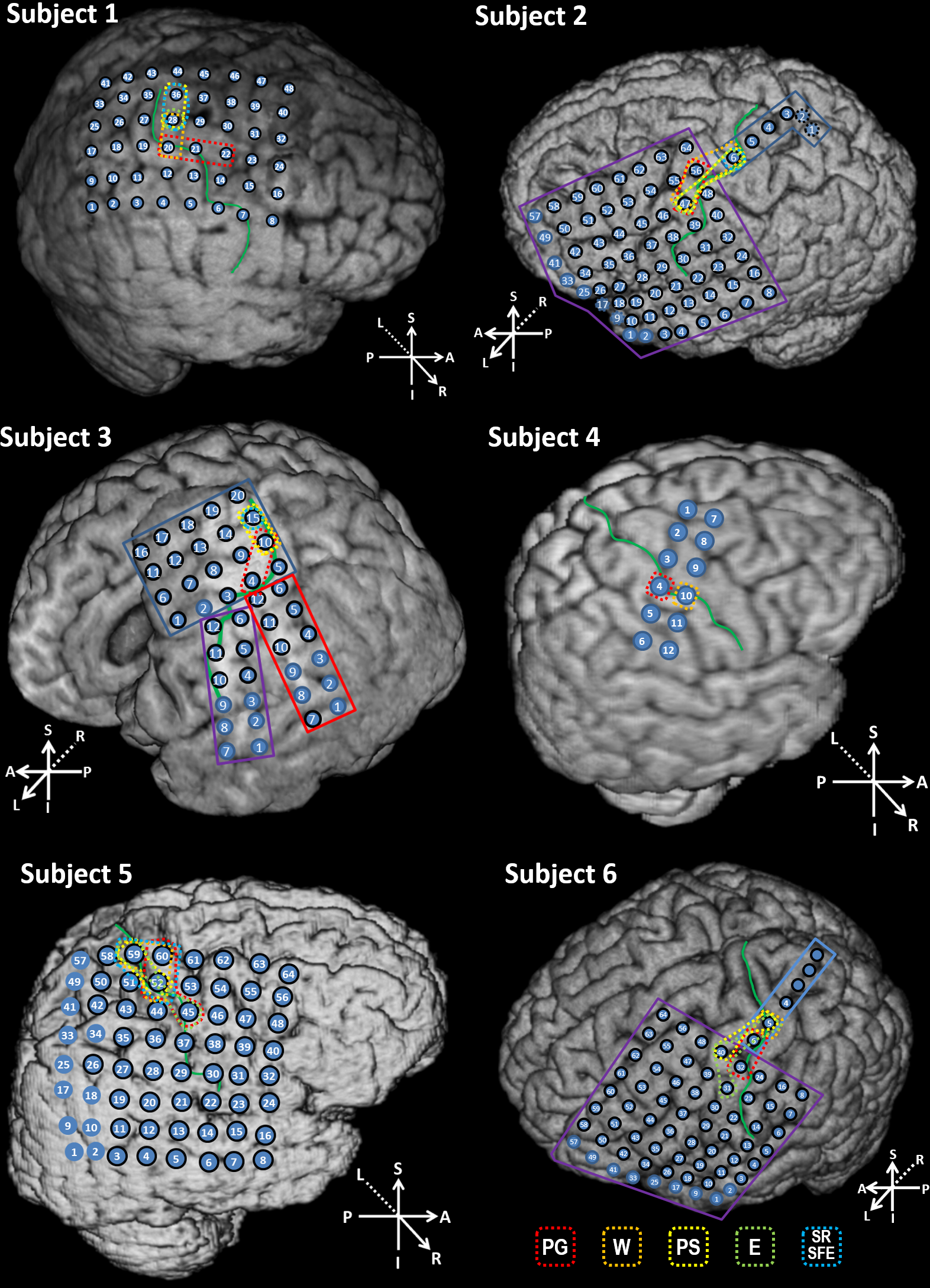}
\caption{
Magnetic resonance imaging (MRI) and computed tomography (CT) co-registration of ECoG electrodes locations. The colored circles represent electrodes, and the black outlined circles are electrodes that were recorded in the present study (limited by amplifier capacity). The solid green lines delineate the central sulcus. Colored dotted lines denote the electrodes with the highest correlation to the corresponding movement types (based on Table~\ref{tab:intermittent} for Subjects 1-5 and Table~\ref{tab:continuous} for Subject 6). Electrode localization was performed using the technique described in \cite{ptwang:13}. Refer to Table~\ref{tab:demographics} for descriptions of electrode locations.
}
\label{fig:grids}
\end{figure*}

\begin{table*}[!htpb]
\caption{
Cross-correlation results and spectral energy of the top channels for all subjects during intermittent movement. 
F, I, and E denote flexion, idle, and extension segments, respectively. 
Lag time in seconds are provided in parentheses. 
Idle results are calculated at zero lag. 
Flexion/Extension Energy are obtained by integrating the areas under the corresponding high-$\gamma$ power bursts, reported in z-scores $\times$ time.
P-values are from two-tail t-test between the flexion and extension energies.
\fsj{SJ0082} did not perform any intermittent movements.
}
\label{tab:intermittent}
\centering
\begin{tabular}{cclrlrrcl}
  &  Electrode  &  $\rho(P_F, \dot\theta_F)$  &  $\rho(P_I, \dot\theta_I)$  &  $\rho(P_E, \dot\theta_E)$  &  Flex Energy  &  Ext Energy  &  Stronger  &  P-value  \\
\midrule
\multicolumn{9}{l}{\fsj{SJ0084}}    \\
    PG &    G20 &   0.75 (0.02) & -0.21 &  -0.64 (0.01) & 11.28$\pm$3.66 & 14.88$\pm$4.33 & E & 0.0042 \\
       &    G22 &   0.66 (0.02) & -0.29 & -0.42 (-0.06) & 8.48$\pm$3.71 & 5.60$\pm$2.13 & F & 0.0029 \\
     W &    G20 &   0.73 (0.44) & -0.17 & -0.55 (-0.11) & 14.23$\pm$2.50 & 11.18$\pm$3.06 & F & 0.0012 \\
       &    G28 &  0.56 (-0.11) & -0.00 & -0.58 (-0.03) & 7.33$\pm$3.12 & 6.85$\pm$2.06 & - & 0.57 \\
    PS &    G28 &   0.47 (0.00) &  0.03 & -0.62 (-0.18) & 11.81$\pm$2.96 & 9.63$\pm$2.80 & F & 0.014 \\
       &    G36 &   0.58 (0.43) &  0.12 & -0.45 (-0.04) & 8.32$\pm$3.30 & 3.75$\pm$2.02 & F & 1.06$\times 10^{-6}$ \\
     E &    G28 &   0.36 (0.18) & -0.07 & -0.38 (-0.56) & 6.33$\pm$4.29 & 6.16$\pm$3.12 & - & 0.87 \\
    SR &    G28 &   0.45 (0.45) &  0.10 & -0.47 (-0.30) & 11.93$\pm$2.97 & 9.84$\pm$2.59 & F & 0.019 \\
       &    G36 &   0.32 (0.50) &  0.06 & -0.39 (-0.24) & 3.82$\pm$2.00 & 4.32$\pm$1.90 & - & 0.42 \\
   SFE &    G28 &  0.60 (-0.52) &  0.05 &  -0.83 (0.27) & 3.12$\pm$1.98 & 6.46$\pm$1.36 & E & 1.77$\times 10^{-8}$ \\
       &    G36 &  0.56 (-0.41) &  0.08 &  -0.47 (0.20) & 4.72$\pm$1.58 & 5.89$\pm$2.06 & E & 0.035 \\
\midrule
\multicolumn{9}{l}{\fsj{SJ0087}}    \\
    PG &    G56 &   0.76 (0.07) & -0.10 &  -0.75 (0.04) & 9.27$\pm$3.59 & 5.52$\pm$2.06 & F & 8.13$\times 10^{-6}$ \\
       &    G47 &   0.49 (0.07) &  0.03 &  -0.50 (0.10) & 5.19$\pm$3.27 & 6.47$\pm$2.89 & - & 0.12 \\
     W &     S6 &   0.71 (0.09) &  0.18 &  -0.65 (0.07) & 12.75$\pm$5.01 & 15.10$\pm$6.74 & - & 0.16 \\
       &    G47 &   0.42 (0.02) &  0.07 & -0.41 (-0.02) & 5.94$\pm$3.92 & 4.89$\pm$2.16 & - & 0.25 \\
       &    G56 &   0.61 (0.08) &  0.02 & -0.30 (-0.08) & 3.93$\pm$1.73 & 1.88$\pm$2.00 & F & 0.00024 \\
    PS &     S6 &   0.61 (0.08) &  0.05 &  -0.60 (0.07) & 17.46$\pm$8.10 & 12.66$\pm$6.70 & F & 0.019 \\
       &    G47 &  0.40 (-0.05) &  0.05 &  -0.58 (0.00) & 4.55$\pm$2.52 & 6.60$\pm$2.31 & E & 0.0025 \\
     E &     S6 &  0.44 (-0.16) & -0.37 &  -0.59 (0.13) & 26.94$\pm$8.84 & 24.90$\pm$5.83 & - & 0.34 \\
       &    G47 &  0.45 (-0.28) & -0.21 &  -0.34 (0.05) & 12.85$\pm$5.31 & 11.35$\pm$5.69 & - & 0.34 \\
    SR &     S6 &   0.51 (0.18) &  0.08 & -0.32 (-0.34) & 13.67$\pm$3.21 & 10.87$\pm$1.96 & F & 0.0013 \\
   SFE &     S6 &  0.75 (-0.21) & -0.20 &  -0.55 (0.23) & 21.62$\pm$5.57 & 20.70$\pm$5.28 & - & 0.55 \\
\midrule
\multicolumn{9}{l}{\fsj{SJ0088}}    \\
    PG &   LFG4 &   0.62 (0.15) &  0.06 & -0.81 (-0.09) & 5.80$\pm$2.51 & 1.75$\pm$0.98 & F & 1.21$\times 10^{-8}$ \\
       &  LFG10 &  0.54 (-0.05) &  0.06 & -0.62 (-0.10) & 3.67$\pm$2.10 & 3.31$\pm$2.83 & - & 0.64 \\
     W &  LFG10 &  0.61 (-0.19) &  0.15 & -0.44 (-0.24) & 6.87$\pm$3.88 & 11.22$\pm$6.10 & E & 0.0037 \\
       &  LFG15 &  0.71 (-0.21) &  0.19 & -0.48 (-0.23) & 3.77$\pm$1.77 & 3.23$\pm$2.77 & - & 0.43 \\
    PS &  LFG15 &  0.51 (-0.16) &  0.04 & -0.48 (-0.21) & 8.88$\pm$5.84 & 3.77$\pm$2.71 & F & 0.0001 \\
       &  LFG10 &  0.47 (-0.22) & -0.01 & -0.42 (-0.11) & 2.84$\pm$2.56 & 4.25$\pm$3.03 & - & 0.063 \\
     E &  LFG15 &  0.61 (-0.51) &  0.05 & -0.75 (-0.31) & 5.96$\pm$3.11 & 8.88$\pm$3.16 & E & 0.0015 \\
    SR &  LFG15 &  0.48 (-0.42) & -0.12 & -0.61 (-0.41) & 7.89$\pm$2.58 & 6.68$\pm$2.75 & - & 0.11 \\
   SFE &  LFG15 &  0.71 (-0.52) &  0.01 & -0.74 (-0.44) & 5.20$\pm$1.99 & 9.31$\pm$2.32 & E & 1.03$\times 10^{-8}$ \\
\midrule
\multicolumn{9}{l}{\fsj{SJ0085}}    \\
    PG &     F4 &   0.64 (0.32) & -0.05 & -0.42 (-0.05) & 8.66$\pm$4.09 & 2.75$\pm$1.92 & F & 1.71$\times 10^{-8}$ \\
     W &    F10 &  0.39 (-0.51) & -0.04 & -0.26 (-0.59) & 2.53$\pm$1.61 & 2.22$\pm$1.32 & - & 0.5 \\
\midrule
\multicolumn{9}{l}{\fsj{SJ0096}}    \\
    PG &    G52 &   0.65 (0.17) & -0.12 & -0.66 (-0.08) & 12.62$\pm$4.77 & 7.85$\pm$3.63 & F & 0.00023 \\
       &    G60 &   0.64 (0.18) & -0.23 & -0.39 (-0.10) & 14.26$\pm$4.96 & 5.53$\pm$3.17 & F & 1.66$\times 10^{-9}$ \\
       &    G45 &   0.43 (0.17) &  0.03 & -0.42 (-0.01) & 4.07$\pm$3.40 & 3.18$\pm$2.98 & - & 0.33 \\
     W &    G59 &   0.63 (0.02) &  0.08 & -0.52 (-0.07) & 5.96$\pm$2.86 & 3.93$\pm$1.81 & F & 0.0082 \\
       &    G52 &   0.62 (0.13) & -0.07 & -0.45 (-0.11) & 11.18$\pm$4.43 & 6.40$\pm$3.78 & F & 0.00047 \\
       &    G60 &   0.57 (0.10) & -0.09 & -0.56 (-0.11) & 6.59$\pm$1.87 & 2.14$\pm$1.49 & F & 9.55$\times 10^{-11}$ \\
    PS &    G59 &   0.60 (0.00) &  0.07 & -0.55 (-0.06) & 7.28$\pm$4.45 & 6.08$\pm$2.42 & - & 0.26 \\
       &    G52 &   0.54 (0.00) &  0.02 &  -0.50 (0.15) & 7.27$\pm$3.56 & 10.70$\pm$4.73 & E & 0.0089 \\
     E &    G59 &  0.44 (-0.36) &  0.03 & -0.32 (-0.03) & 3.64$\pm$1.99 & 2.68$\pm$1.90 & - & 0.1 \\
       &    G45 &  0.43 (-0.07) & -0.01 &  -0.38 (0.13) & 3.46$\pm$2.14 & 3.36$\pm$2.61 & - & 0.89 \\
    SR &    G59 &  0.41 (-0.51) & -0.05 & -0.39 (-0.44) & 5.01$\pm$3.57 & 3.66$\pm$3.10 & - & 0.17 \\
       &    G60 &  0.40 (-0.50) & -0.01 & -0.24 (-0.39) & 1.64$\pm$1.19 & 1.64$\pm$1.36 & - & 0.99 \\
   SFE &    G59 &  0.50 (-0.53) & -0.02 &  -0.65 (0.33) & 2.10$\pm$2.11 & 2.71$\pm$2.10 & - & 0.28 \\
       &    G52 &  0.38 (-0.52) &  0.09 &  -0.58 (0.40) & 0.00$\pm$0.00 & 2.23$\pm$1.22 & E & 3.11$\times 10^{-11}$ \\
\bottomrule
\end{tabular}
\end{table*}

\begin{table}[!htbp]
\caption{
Top correlating ($\rho$) electrodes and visually distinguishable electrodes for all Subjects during continuous movements. Individual electrode details in Appendix Table~\ref{atab:continuous}.
}
\label{tab:continuous}
\centering
\begin{tabular}{cccl}
\toprule
Subject  &  Movement  &  Highest $\rho$  &  Top Channels  \\
\midrule
\sj{SJ0084}  &    PG  &  0.57  &       G14, G22, G21  \\
\sj{SJ0084}  &     W  &  0.21  &            G20, G28  \\
\sj{SJ0084}  &    PS  &  0.07  &                 G28  \\
\sj{SJ0084}  &     E  &  0.18  &                 G28  \\
\sj{SJ0084}  &    SR  &  0.47  &       G20, G28, G36  \\
\sj{SJ0084}  &   SFE  &  0.47  &       G28, G22, G29  \\
\midrule
\sj{SJ0087}  &    PG  &  0.73  &        G64, G56, S6  \\
\sj{SJ0087}  &     W  &  0.67  &        G56, G64, S6  \\
\sj{SJ0087}  &    PS  &  0.11  &             S6, G47  \\
\sj{SJ0087}  &     E  &  0.21  &        S6, G47, G56  \\
\sj{SJ0087}  &    SR  &  0.27  &        G64, G39, S6  \\
\sj{SJ0087}  &   SFE  &  0.44  &        S6, G39, G56  \\
\midrule
\sj{SJ0088}  &    PG  &  0.64  &   LPS12, LPS6, LFG4  \\
\sj{SJ0088}  &     W  &  0.29  &  LFG4, LFG15, LFG10  \\
\sj{SJ0088}  &    PS  &  0.26  &  LFG5, LFG10, LFG15  \\
\sj{SJ0088}  &     E  &  0.17  &        LFG15, LFG10  \\
\sj{SJ0088}  &    SR  &  0.26  &        LFG15, LFG10  \\
\sj{SJ0088}  &   SFE  &  0.44  &  LFG4, LFG10, LFG15  \\
\midrule
\sj{SJ0085}  &    PG  &  0.73  &         F4, F10, F3  \\
\sj{SJ0085}  &     W  &  0.29  &                 F10  \\
\midrule
\sj{SJ0096}  &    PG  &  0.58  &       G60, G37, G52  \\
\sj{SJ0096}  &     W  &  0.29  &       G52, G60, G59  \\
\sj{SJ0096}  &    PS  &  0.28  &       G52, G60, G59  \\
\sj{SJ0096}  &     E  &  0.19  &                 G59  \\
\sj{SJ0096}  &    SR  &  0.37  &            G39, G59  \\
\sj{SJ0096}  &   SFE  &  0.41  &       G59, G52, G60  \\
\midrule
\sj{SJ0082}  &    PG  &  0.58  &       PS5, PS6, G32  \\
\sj{SJ0082}  &     W  &  0.39  &                 PS5  \\
\sj{SJ0082}  &    PS  &  0.12  &                 PS5  \\
\sj{SJ0082}  &     E  &  0.62  &       PS5, G31, G40  \\
\bottomrule
\end{tabular} 
\end{table}

\begin{figure*}[!ht]
\centering
\includegraphics[width=\linewidth]{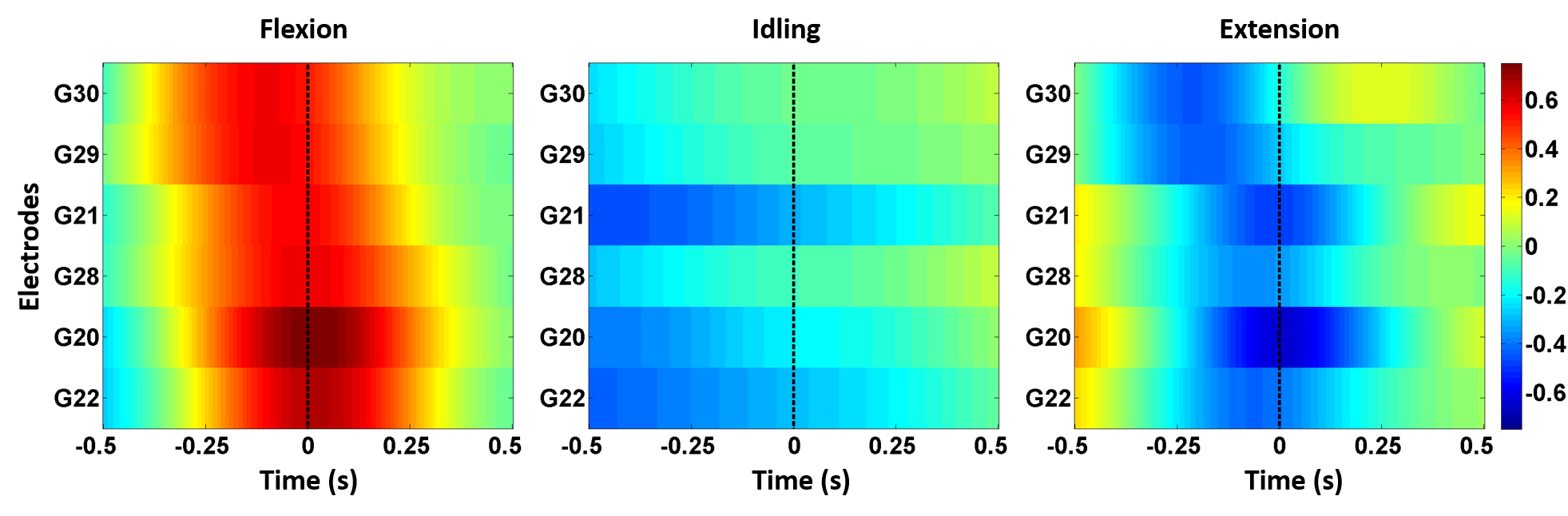}
\centering
\includegraphics[width=1.0\linewidth]{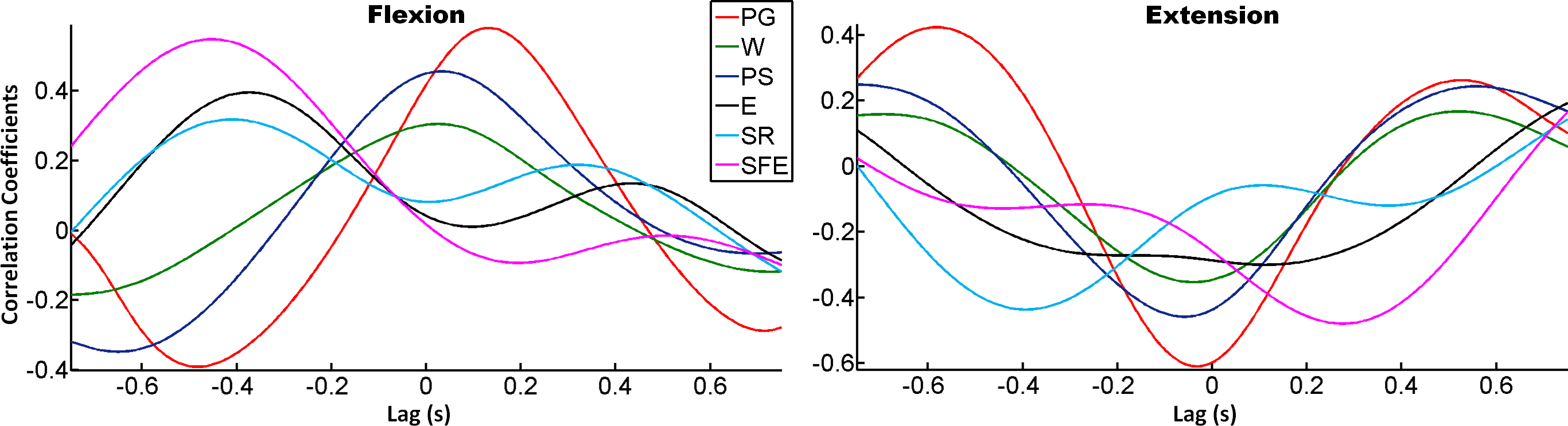}
\caption{
(Top) Representative lag-correlation diagrams of \fsj{SJ0084} for PG intermittent movement. 
For each electrode, the colors represent the cross-correlation at various lag times. 
Flexion, idling, and extension have their own lag-correlation diagram.
Note how correlations strengthen before the onset of movement (time=0~s).
The supplementary motor area electrodes, G29 and G30 (see Fig.~\ref{fig:grids}), have correlations that strengthen before the motor cortex electrodes, G20, G21, G22, and G28.
During idling, correlations are weak.
(Bottom) Correlation vs. lag times averaged across subjects for each intermittent movement type. 
Arm supination and shoulder internal rotation are also grouped with flexion.
}
\label{fig:subject1lagcorrelation}
\end{figure*}

\begin{table}[]
\caption{
Ratio of the number of identified $P(t)$ peaks to total number of movement cycles in small and large joint movements.
}
\label{tab:numidentified}
\centering
\begin{tabular}{ccccc}
\toprule
       & \multicolumn{2}{c}{Continuous} & \multicolumn{2}{c}{Intermittent} \\
Movement  &  Flexion  &  Extension  &  Flexion  &  Extension  \\
\midrule
   PG  &  92\%  &  17\%  &  98\%  &  88\%  \\
    W  &  86\%  &  60\%  &  94\%  &  96\%  \\
   PS  &  84\%  &  77\%  &  98\%  &  99\%  \\
    E  &  92\%  &  82\%  &  97\%  &  97\%  \\
   SR  &  86\%  &  94\%  &  93\%  &  91\%  \\
  SFE  &  92\%  &  90\%  &  74\%  &  89\%  \\
\midrule
Small  &  88\%  &  44\%  &  97\%  &  94\%  \\
Large  &  90\%  &  88\%  &  88\%  &  92\%  \\
\bottomrule
\end{tabular}
\end{table}

\subsection{Spatial Patterns}

\subsubsection{M1}
Based on the results from the intermittent and continuous movement paradigm, the electrodes over M1 with the strongest correlation for each movement type were encircled with dotted lines in Fig.~\ref{fig:grids} to characterize their spatial distribution. 
These electrode groups are more lateral for movements at distal joints (PG, W, and PS), and are progressively more medial for more proximal joints (E, SR, and SFE). However, there is a significant amount of overlap between each of these groups. 
Of note, within each movement type, the power bursts associated with both directions (e.g. flexion and extension) were present on the same electrodes. There were no instances where an electrode was found to be correlated to flexion and not to extension (and vice versa).

\subsubsection{Non-M1 Areas}
\begin{figure}[!ht]
\centering
\includegraphics[width=1.0\linewidth]{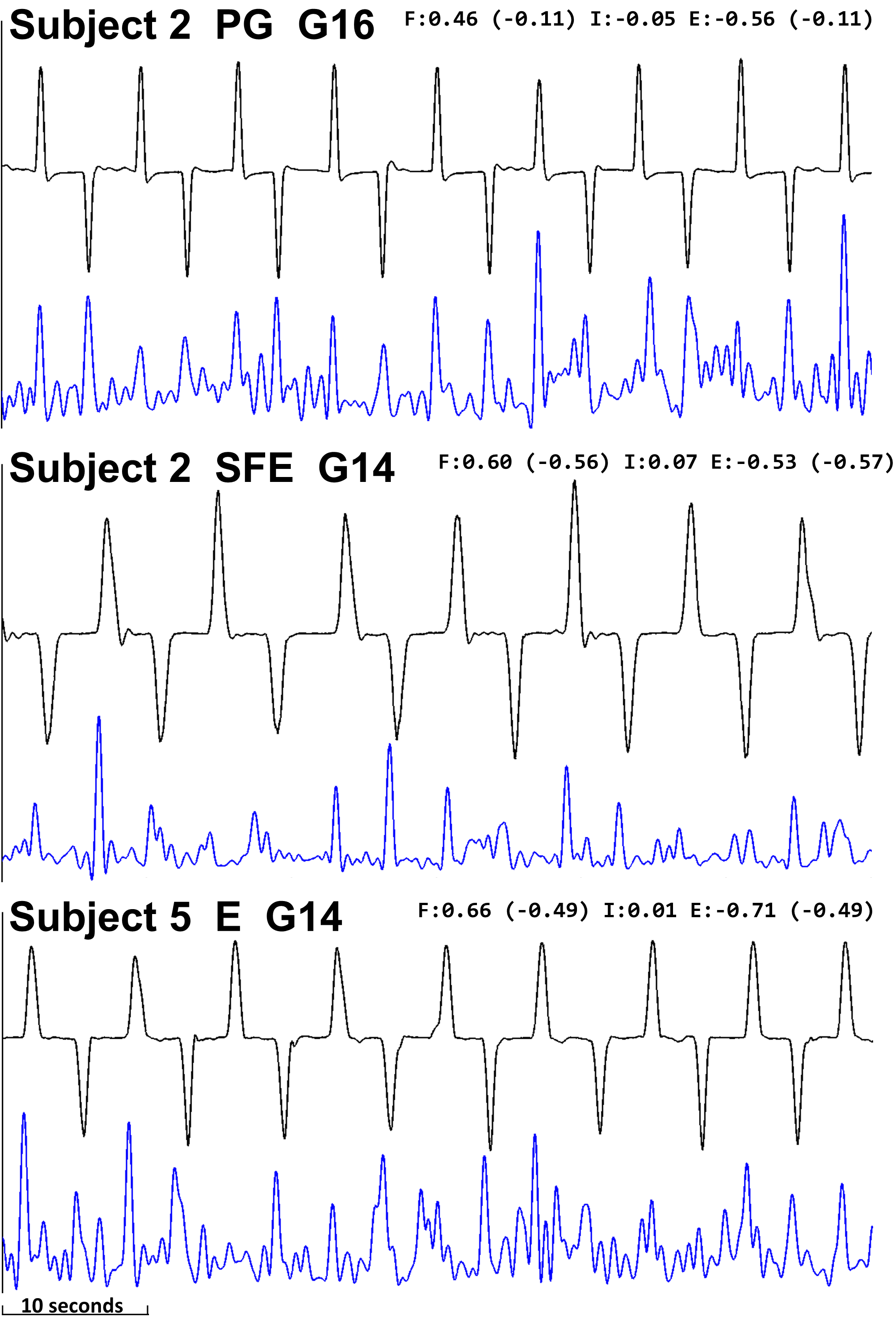}
\caption{
A representative segment of $P(t)$ (blue trace) and corresponding $\dot\theta(t)$ (black) at selected auditory cortex electrodes. At the top-right corner of each panel are the correlations and optimal lags (seconds, in parentheses) for flexion (F) and extension (E) epochs. The correlation for idling (I) was calculated without lag optimization. 
}
\label{fig:auditoryintermittent}
\end{figure}

In an ancillary, yet important observation, movement-related high-$\gamma$ power bursts were observed in electrodes covering other areas outside of M1, such as the supplementary motor area, posterior parietal cortex, and auditory cortex. 
For example, Subjects \nsj{SJ0087} and \nsj{SJ0096} exhibited highly correlated power bursts in the auditory cortex (see Fig.~\ref{fig:auditoryintermittent}), similar in shape to those found in M1 electrodes. These power bursts may be the result of subjects receiving verbal cues on every flexion and extension epochs. 

\section{Discussion}

The pioneering studies in \cite{crone:98} first identified that there is an event related synchronization (ERS) in the ECoG high-$\gamma$ band associated with cortical activation. 
In addition, the studies \cite{crone:98,pfurtscheller:03,miller:07b,tyanagisawa:12} demonstrated that a crude form of motor somatotopy can be observed in the representation of high-$\gamma$ ERS (e.g. mouth/tongue related tasks were represented laterally on M1, while arm and leg motor tasks were represented in progressively more medial locations). Although spectro-spatial events have been well described, there have been no rigorous physiological research efforts to further characterize how the high-$\gamma$ may relate to kinematic parameters of movement. Nevertheless, many BCI studies have subsequently attempted to utilize high-$\gamma$ band information to build decoders for movement kinematics without fully understanding the ECoG motor physiology and its limitations. As a result, it is not surprising that the trajectory decoding results are generally poor to moderate at best. 
 
The current study is a unique, comprehensive study which examines the high-$\gamma$ characteristics in M1 across 6 upper extremity movements in order to provide greater details on how the brain encodes for upper extremity movements. The results provide insight into the spatial and temporal relationship between high-$\gamma$ and movement kinematics, add to the understanding of how the brain encodes for upper extremity movements, and may define the signal features as well as potential limits which ECoG signals can be exploited for BCI applications.

\subsection{High-$\gamma$ Encoding and Spatial Distribution}

During the intermittent motor task, the visual similarity between $P(t)$ and $\dot\theta(t)$, as well as the high positive (negative) cross-correlation values during flexion (extension) epochs suggests that ECoG high-$\gamma$ power strongly encodes for elementary upper extremity movement velocity. 
Conversely, idling periods are characterized by a lack of correlation, and the $P(t)$ signal appears desynchronized (lower amplitude, noisy, and chaotic). 
Table \ref{tab:intermittent} indicates that this encoding pattern is present for all elementary movements in relevant M1 electrode(s). 
 The high-$\gamma$ bursts are distributed in a manner consistent with classical somatotopy (distal joints located more laterally in M1, while proximal joints are located more medially). 
These findings point to the existence of somatotopically arranged neuronal generators that drive each movement type.
When active, these generators appear to behave in a similar spectral-temporal manner by producing high-$\gamma$ bursts. 
Furthermore, within each movement type, the observed energy disparity between power bursts associated  with flexion and extension is likely not due to a difference in the amount of movement during each direction (since both movements are over the full range of motion). 
Instead, this implies the presence of spatially separate neuronal generators for each movement direction that are differentially sensed by each electrode.  
However, given the significant spatial overlap between the relevant electrodes for each movement type and direction, it will be difficult to distinguish between each of them.
Such a problem may require the use of more sophisticated signal processing techniques (e.g. source localization and separation) and/or signals from ECoG grids with higher electrode density to disentangle these neuronal generators.

This high-$\gamma$ encoding for movement velocity agrees with the prior work by Anderson et al. \cite{nranderson:12}. However, \cite{nranderson:12} only examined the relationship between high-$\gamma$ and movement velocity during a complex center out, and circle tracing task. Hence, further deconstruction for mapping and characterization of the individual upper extremity elementary movements was not possible. 
The ECoG high-$\gamma$ motor maps from Crone et al. (tongue, arm, foot) \cite{crone:98}, Pfurtscheller et al. (fingers and tongue) \cite{pfurtscheller:03}, Miller et al. (tongue, hand, shoulder, hips) \cite{miller:07b}, and Yanagisawa et al. (hand and elbow) \cite{tyanagisawa:12} provided a crude sense of somatatopic organization in M1. The current study provides the first detailed mapping of the major upper extremity degrees-of-freedom, and reinforces the concept of somatotopic arrangement within M1. The findings are consistent the fMRI findings of \cite{jdmeier:08}, where there are somtatopically arranged, but highly-overlapped representation areas for finger, wrist, forearm, and elbow movements. Although it is possible to interpret this overlap of the relevant ECoG electrodes as being caused by the interspersing of responsible neuronal populations amongst one another, an alternative explanation is that the volume conduction from each neuronal generator causes overlapping signals when the electrode resolution is still inadequate.
Penfield and Boldrey~\cite{wpenfield:37} showed that electrical stimulation on different pre-central areas can elicit isolated flexion or extension for an individual upper extremity joint (ranging from the fingers to the shoulder), suggesting the presence of spatially separate brain areas for each of the joints and movement directions. However, based on their figures, the spatial separation seems to be in the order of millimeters, which may explain why the significant degree of overlap seen here with standard ECoG grids.

\subsection{Non-M1 Areas}
The presence of power bursts in areas outside of M1 implies that high-$\gamma$ power bursts are not exclusive to M1, and that other brain areas subserving non-motor functions exhibit similar high-$\gamma$ bursts when activated, which is consistent with \cite{crone:01}. This is a confounding factor that can complicate a BCI training data collection, as verbal cues and other stimuli correlated to movement can inadvertently be extracted as features encoding for movement trajectories, resulting in a systematic error. To mitigate this confounding factor, input channel selection should be based on anatomy, or training data collection should be self paced. 
On the other hand, it may also indicate that areas such as premotor and supplementary motor can be further studied to elucidate how they encode movements, which may help increase BCI performances in the future. However, additional analysis and discussion upon how these areas temporally relate to M1 or are networked with one another is beyond the scope of the present study.

\subsection{Small and Large Joints}
\begin{figure}
\centering
\includegraphics[width=0.9\linewidth]{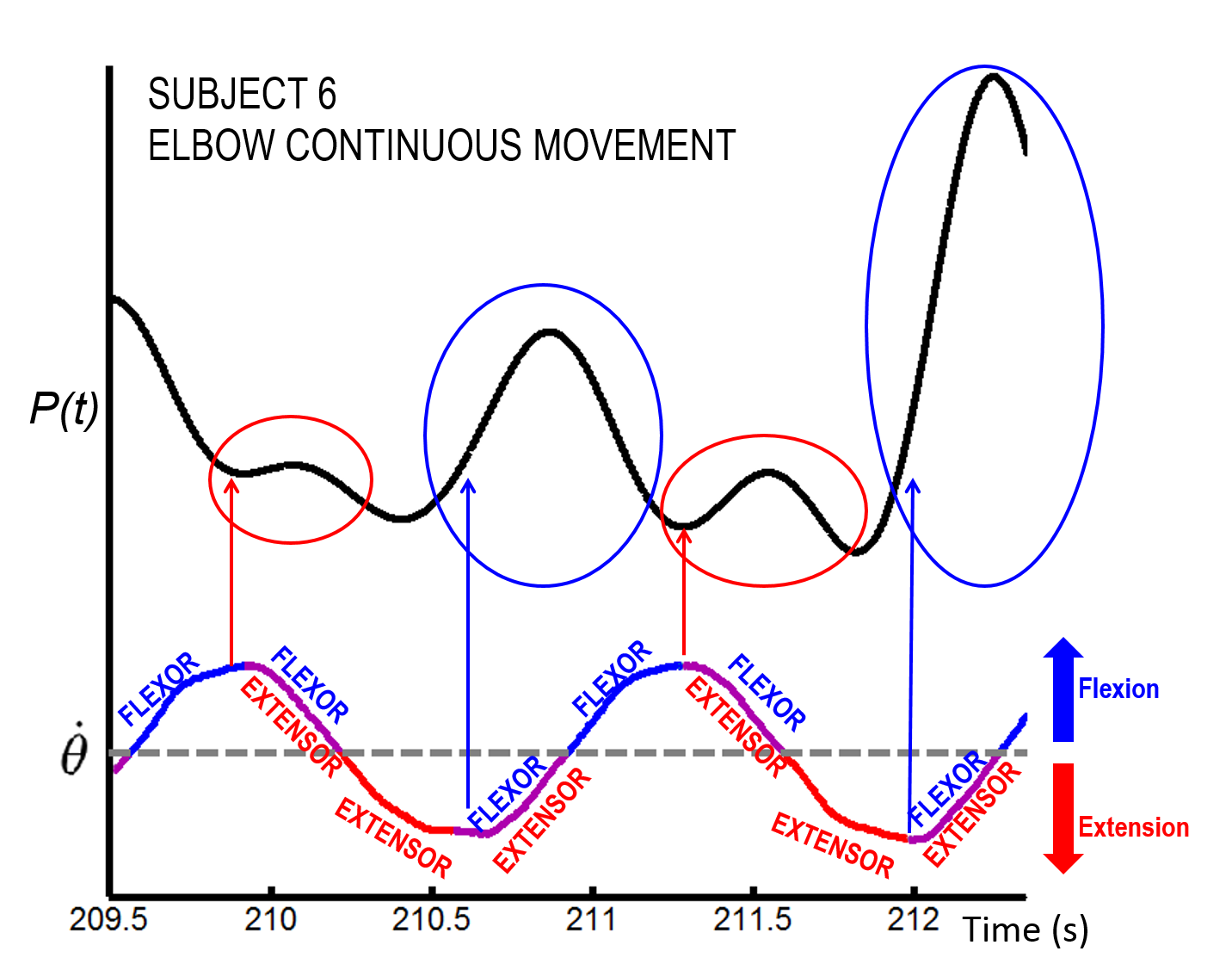}
\caption{
Closer examination of the early $P(t)$ peaks on a large joint continuous movement. 
At the times indicated by the red arrows, eccentric extensor muscle activation at the maximum flexion velocity serves to dampen the flexion movements. 
At the time indicated by the blue arrows, eccentric flexor muscle activation at the maximum extension velocity serves to decelerate the extension movements.
Note that the grey dotted line indicates when velocity is 0$^{\circ}$/s.
}
\label{fig:bigjointexplain}
\end{figure}

\begin{figure*}[!ht]
\centering
\includegraphics[width=1.0\linewidth]{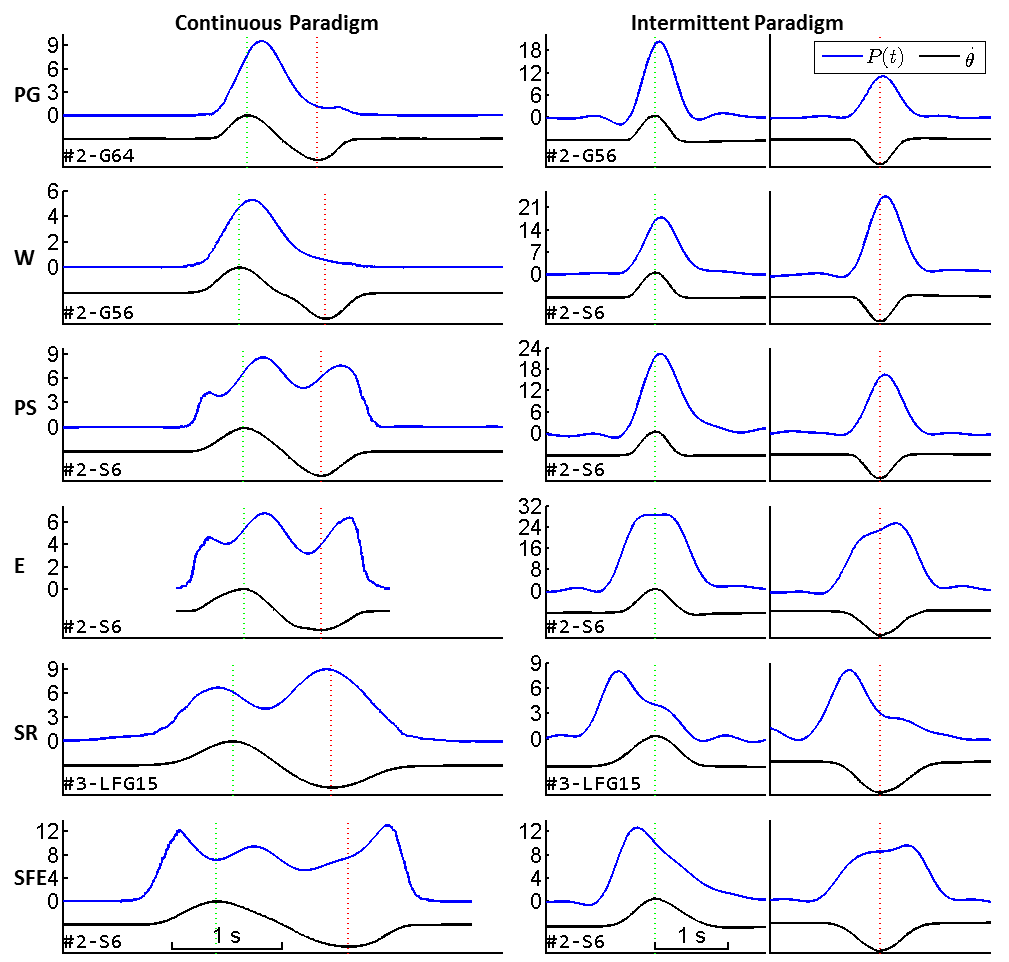}
\caption{
$P(t)$ and $\dot{\theta}$ signals averaged across individual flexion and extension cycles. 
Blue lines: $P(t)$, black lines: $\dot{\theta}$.
Subject numbers (\#) and electrodes are shown at the bottom-left corner of each plot.
The amplitudes of $P(t)$ are standardized to the MAD of the idle epochs.
For continuous movements, time zero is set at the middle between a flexion and the next extension.
For intermittent movements, the idling periods between flexions and extensions are not shown.
Note that for the continuous paradigm, $P(t)$ signals from the next cycle may wrap-around.
}
\label{afig:amp}
\end{figure*}

As shown in Figs. \ref{fig:phaseshift2} and \ref{afig:amp}, a leading phase shift of $\sim$1/4-cycle was observed in large joint continuous movements, whereas no significant phase shift was observed in small joint movements. The phase shift in large joint continuous movements may be due to the difference in the kinematic profile between small and large joint movements. 
In small joint movements, the relatively lower mass of the appendage (e.g. fingers, or hand) requires little activity from the opposing muscle group to transition to the opposite direction of movement. A person can simply allow the limits in range of motion to stop the movement before executing the opposite movement.
However, in order to achieve seamless continuous movements in larger joints, it is well known that the increased inertia necessitates eccentric contraction of the opposing muscle group for braking and transitioning to movement in the opposite direction.
More specifically, eccentric activation of the opposing muscle groups are known to initiate at the maximal flexion and extension velocities \cite{flestienne:79}. 
For example, during large joint movements such as SFE, subjects would utilize anterior deltoid muscles to lift the entire arm against gravity for the flexion phase. This is followed by the eccentric contraction of the posterior deltoids to dampen the flexion and transition to extension movement.
In addition, due to the higher inertia, muscle activation in large joints are expected to precede actual movement in order to generate adequate force and tension. Kinematically, this would result in a 1/4-cycle leading eccentric activation of the opposing muscle group as depicted in Fig.~\ref{fig:bigjointexplain}.
Since the high- $\gamma$ bursts coincide with this pattern, these observations give rise to a possibility that the high-$\gamma$ band power more directly encodes for muscle activity. 
Note that during large joint intermittent movements, some braking may also occur, and may explain why binotched high-$\gamma$ power bursts are frequently seen (see Figs.~\ref{fig:intermittentwaveform} and \ref{afig:amp} for movements E and SFE). However, the rapid transition between flexion and extension is not required, and hence the phase shift is not as prominent when compared to continuous movements (Fig.~\ref{fig:phaseshift2}). 
These observations serve to further support this hypothesis. Ultimately, since the current study did not examine electromyogram (EMG) data or systematically vary the amount of force exerted during the movements, this will require additional formal study to confirm. 

\subsection{Correlations in the Continuous Paradigm}
Correlation coefficients for the continuous paradigm are considerably lower than those for the intermittent paradigm due to a combination of physiological and mathematical causes. 
From the physiological standpoint, the disappearance of the expected extension associated high-$\gamma$ power burst during fast movements, and their re-emergence whenever subjects introduce a brief pause as they transition between flexion and extension (Fig.~\ref{fig:bigsmalljoint}a), suggests a merging of flexion and extension related power bursts during fast, continuous movements.
This is usually only observed in PG movements. Mathematically, since the $P(t)$ waveform better matches the $\dot{\theta}$ waveform, the cross-correlation for continuous PG movements tend to be the highest (seen in 5 out of 6 subjects, Table~\ref{tab:continuous}).
However, these correlations are still weaker than their intermittent counterparts. This is likely due to the distortion caused by the merging of flexion and extension power bursts. 
For the remaining joints, their increased size generally precludes performing flexion and extension movements in rapid succession. As a result, power bursts are visible for both flexion and extension epochs, similar to intermittent movements (see Fig.~\ref{fig:bigsmalljoint}b and c, and Fig.~\ref{afig:amp}).
Since a high-$\gamma$ power burst during extension causes anti-correlation, it reduces the overall correlation for the entire movement session as seen in movements other than PG in Table~\ref{tab:continuous}. 
These findings have many implications for ECoG-based BCIs.
The merging of the power bursts during rapid alternation of flexion and extension illustrates the lack of spatial resolution in standard ECoG grids.
As above, this also indicates that source localization and separation techniques and higher resolution signals may help to separate the activities underlying flexion and extension.
Also, the moderate-to-poor performances in other ECoG-based decoders (\cite{kjmiller:09,jkubanek:09,nliang:12,zwang:11,sacharya:10,hlbenz:12,gschalk:07,tpistohl:08,jcsanchez:08}) are due to the use of continuous movements in training, which do not generalize to all manners of movements.
Finally, due to the nature of $P(t)$ during continuous movement, measures such as correlation coefficient may not be the best method to objectively quantify how well ECoG signals encode for movement trajectory in the continuous paradigm. 
Instead, different quantitative approaches such as mutual information may need to be used for this purpose.

\subsection{Additional Questions, Hypotheses, and Future Directions}
Based on the observations of the current study and of prior reports from the literature, there is evidence to suggest that M1 contains separate neuronal generators that are responsible for encoding movements on an individual muscle group basis (thereby controlling the movement velocity and direction at each joint).
In turn, it can also be hypothesized that complex movements (e.g. reaching tasks) are produced by the summation of multiple generators acting simultaneously. Although current literature for both neuronal single cell populations \cite{georgopoulos:82} and ECoG (\cite{nranderson:12,jcsanchez:08} propose that M1 may encode for arm trajectory (in a directional tuning model), these studies only utilized a center-out reaching task, and did not consider what the underlying patterns of individual muscle groups are. For example, right arm movement to the 0$^{\circ}$ will always be associated with preferential firing of a population of neurons or a specific ECoG pattern. However, this type of movement always activates the infraspinatus (shoulder external rotation) and triceps (elbow extension), and without a control experiment to determine if the neurons which are preferentially firing actually consist of 2 subpopulations (one for each muscle). Likewise, it is unknown if ECoG potentials from 2 different neuronal generators occurred simultaneously. 

It also is unclear if the high-$\gamma$ bursts are simply present with the onset of movement, or if they persist for the duration of the movement. It is also unclear if and how they vary with different velocity profiles. For example, a single power burst at the onset of movement may encode for both the velocity and duration of movement, while other centers of the nervous system may be responsible for transducing these signals into muscle activity.
Based on the current study, the duration of the high-$\gamma$ bursts appeared to follow the movement duration. However, the duration was not systematically varied to establish this relationship. In a similar manner, the velocity profile of movement was not systematically varied. These relationship between high-$\gamma$ warrant additional study as it will help provide a more comprehensive understanding of brain motor control for the upper extremities.

\section{Conclusion}
Despite minimal processing, the correlation between $P(t)$ and $\dot\theta(t)$ at a single channel is already as high as (and occasionally higher than) the decoded results reported in the prior literature \cite{gschalk:07, kjmiller:09, jkubanek:09, nliang:12, zwang:11, sacharya:10, hlbenz:12, tpistohl:08, jcsanchez:08}. 
Hence, the authors hypothesize that using $P(t)$ from M1 as an input feature for future BCI decoding algorithms may significantly boost the decoding accuracies. 
However, an additional fundamental understanding of ECoG neurophysiology may be necessary before a useful and generalizable model of upper extremity movements can be designed. 
Specifically, it is unclear if further spatial or spectral separation of individual movement types, or flexion and extension generators, is possible. 
Currently, it seems that the same 2-3 M1 channels are involved across all movements in all subjects, indicating that the separate neuronal generators of upper extremity movements are densely packed in a small area of M1, which may make it difficult to resolve them. 
This warrants further investigation to determine how these generators can be better distinguished, and subsequently exploited for BCI control. 
This will require the application of more sophisticated signal processing techniques, or possibly higher resolution signals, such as those from mini- or micro-ECoG grids.

\section{ACKNOWLEDGMENTS}

The authors would like to thank Angelica Nguyen and Christel Jean for their assistance in setting up the experiments.


\bibliography{ecog_biblio}
\bibliographystyle{ieeetr}

\section{APPENDIX}
\begin{table*}[!htbp]
\caption{
Cross-correlation results of the top channels for all Subjects during continuous movements. 
The cross correlation of movement epochs, $\rho_M$, are lag optimized.
The correlations of idle epochs, $\rho_I$, are calculated at the same lag.
$\star$: Visually distinct movement epochs based on $P(t)$ despite poor correlations.
$\dagger$: Visually indistinguishable movement epochs based on $P(t)$ despite good correlations.
}
\label{atab:continuous}
\begin{minipage}[t]{0.5\linewidth}
\begin{tabular}[t]{lclrrr}
\toprule
Subject  &  Movement  &  Electrode  &  $\rho_M$  &  $\rho_I$  &  Lag (s)  \\
\midrule
\sj{SJ0084}  &   PG  &              G14  &  0.57  &  -0.01  &   0.39  \\
             &       &              G22  &  0.54  &  -0.09  &   0.21  \\
             &       &              G21  &  0.53  &   0.01  &   0.36  \\
\cmidrule{2-6}
             &    W  &    G20$^{\star}$  &  0.21  &   0.01  &   0.75  \\
             &       &    G28$^{\star}$  &  0.19  &  -0.16  &  -0.13  \\
\cmidrule{2-6}
             &   PS  &    G28$^{\star}$  &  0.07  &   0.03  &   0.48  \\
\cmidrule{2-6}
             &    E  &    G28$^{\star}$  &  0.18  &   0.25  &   0.11  \\
\cmidrule{2-6}
             &   SR  &              G20  &  0.47  &   0.16  &   0.54  \\
             &       &    G28$^{\star}$  &  0.29  &   0.09  &   0.16  \\
             &       &    G36$^{\star}$  &  0.16  &  -0.05  &  -0.75  \\
\cmidrule{2-6}
             &  SFE  &              G28  &  0.47  &  -0.19  &  -0.75  \\
             &       &              G22  &  0.45  &  -0.12  &  -0.75  \\
             &       &              G29  &  0.45  &  -0.21  &  -0.75  \\
\midrule
\sj{SJ0087}  &   PG  &              G64  &  0.73  &   0.16  &   0.13  \\
             &       &              G56  &  0.68  &   0.10  &   0.03  \\
             &       &               S6  &  0.43  &  -0.01  &   0.05  \\
\cmidrule{2-6}
             &    W  &              G56  &  0.67  &   0.17  &   0.13  \\
             &       &              G64  &  0.49  &   0.03  &   0.17  \\
             &       &               S6  &  0.33  &   0.21  &   0.23  \\
\cmidrule{2-6}
             &   PS  &     S6$^{\star}$  &  0.10  &  -0.01  &   0.60  \\
             &       &    G47$^{\star}$  &  0.11  &   0.00  &   0.45  \\
\cmidrule{2-6}
             &    E  &     S6$^{\star}$  &  0.10  &   0.14  &   0.10  \\
             &       &    G47$^{\star}$  &  0.21  &   0.07  &  -0.23  \\
             &       &    G56$^{\star}$  &  0.13  &  -0.08  &  -0.06  \\
\cmidrule{2-6}
             &   SR  &    G64$^{\star}$  &  0.27  &  -0.01  &   0.27  \\
             &       &    G39$^{\star}$  &  0.25  &   0.19  &   0.37  \\
             &       &     S6$^{\star}$  &  0.22  &   0.28  &   0.34  \\
\cmidrule{2-6}
             &  SFE  &               S6  &  0.44  &  -0.05  &  -0.47  \\
             &       &              G39  &  0.43  &  -0.01  &  -0.34  \\
             &       &              G56  &  0.42  &  -0.12  &  -0.47  \\
\midrule
\sj{SJ0088}  &   PG  &            LPS12  &  0.64  &   0.00  &   0.23  \\
             &       &             LFG4  &  0.56  &  -0.11  &   0.11  \\
             &       &             LFG9  &  0.50  &  -0.06  &   0.09  \\
\cmidrule{2-6}
             &    W  &   LFG4$^{\star}$  &  0.29  &   0.05  &  -0.75  \\
             &       &  LFG15$^{\star}$  &  0.23  &   0.11  &  -0.29  \\
             &       &  LFG10$^{\star}$  &  0.20  &   0.17  &  -0.75  \\
\cmidrule{2-6}
             &   PS  &   LFG5$^{\star}$  &  0.26  &  -0.02  &   0.66  \\
             &       &  LFG10$^{\star}$  &  0.17  &   0.05  &  -0.73  \\
             &       &  LFG15$^{\star}$  &  0.14  &  -0.03  &   0.07  \\
\cmidrule{2-6}
             &    E  &  LFG15$^{\star}$  &  0.17  &  -0.08  &  -0.75  \\
             &       &  LFG10$^{\star}$  &  0.09  &   0.02  &  -0.75  \\
\cmidrule{2-6}
             &   SR  &            LFG15  &  0.26  &   0.03  &   0.75  \\
             &       &            LFG10  &  0.08  &  -0.10  &  -0.75  \\
\cmidrule{2-6}
             &  SFE  &             LFG4  &  0.44  &  -0.05  &  -0.64  \\
             &       &            LFG10  &  0.30  &  -0.03  &  -0.63  \\
             &       &            LFG15  &  0.21  &  -0.18  &  -0.75  \\
\bottomrule
\end{tabular} 
\end{minipage}%
\begin{minipage}[t]{0.5\linewidth}
\centering
\begin{tabular}[t]{lclrrr}
\toprule
Subject  &  Movement  &  Channel  &  $\rho_M$  &  $\rho_I$  &  Lag (s)  \\
\midrule
\sj{SJ0085}  &   PG  &               F4  &  0.73  &   0.03  &   0.28  \\
             &       &              F10  &  0.43  &  -0.04  &   0.34  \\
             &       &               F3  &  0.43  &  -0.06  &   0.25  \\
\cmidrule{2-6}
             &    W  &              F10  &  0.29  &  -0.05  &   0.33  \\
\midrule
\sj{SJ0096}  &   PG  &              G60  &  0.58  &  -0.07  &   0.16  \\
             &       &              G37  &  0.58  &   0.04  &   0.21  \\
             &       &              G52  &  0.50  &   0.00  &   0.19  \\
\cmidrule{2-6}
             &    W  &    G52$^{\star}$  &  0.29  &   0.06  &   0.22  \\
             &       &    G60$^{\star}$  &  0.23  &   0.13  &   0.17  \\
             &       &    G59$^{\star}$  &  0.17  &   0.24  &   0.16  \\
\cmidrule{2-6}
             &   PS  &    G52$^{\star}$  &  0.28  &   0.21  &  -0.41  \\
             &       &    G60$^{\star}$  &  0.27  &   0.18  &  -0.29  \\
             &       &    G59$^{\star}$  &  0.21  &   0.31  &  -0.23  \\
\cmidrule{2-6}
             &    E  &    G59$^{\star}$  &  0.19  &   0.02  &  -0.36  \\
\cmidrule{2-6}
             &   SR  &              G39  &  0.37  &   0.11  &   0.15  \\
             &       &  G23$^{\dagger}$  &  0.36  &   0.03  &   0.34  \\
             &       &    G59$^{\star}$  &  0.17  &  -0.19  &   0.53  \\
\cmidrule{2-6}
             &  SFE  &              G59  &  0.41  &  -0.03  &  -0.48  \\
             &       &              G52  &  0.39  &  -0.02  &  -0.49  \\
             &       &              G60  &  0.38  &  -0.02  &  -0.43  \\
\midrule
\sj{SJ0082}  &   PG  &              PS5  &  0.58  &   0.19  &   0.11  \\
             &       &              PS6  &  0.54  &  -0.05  &   0.06  \\
             &       &              G32  &  0.49  &   0.00  &   0.16  \\
\cmidrule{2-6}
             &    W  &              PS5  &  0.39  &   0.08  &   0.14  \\
\cmidrule{2-6}
             &   PS  &    PS5$^{\star}$  &  0.12  &   0.00  &   0.40  \\
\cmidrule{2-6}
             &    E  &              PS5  &  0.62  &   0.16  &  -0.35  \\
             &       &              G31  &  0.47  &  -0.03  &   0.12  \\
             &       &    G40$^{\star}$  &  0.22  &   0.38  &  -0.29  \\
\bottomrule
\end{tabular}
\end{minipage}
\end{table*}

\begin{sidewaystable*}[]
\caption{Phase-shift analysis of small and large joint movements for both continuous and intermittent paradigms, broken down into individual subjects (see Fig.~\ref{fig:phaseshift2} for summary). Period denotes the time taken to complete one movement cycle in the continuous paradigm. Lags denote the time between the $P(t)$ peaks and the $\dot{\theta}$ peaks and are represented as percentages of the periods for the continuous paradigm and as time for the intermittent paradigm ($P(t)$ leads $\dot{\theta}$ when lag is negative), $\pm$ one standard deviation. Num Flex (Extd) denote the number of identified flexions (extensions) out of the number of total flexion (extensions). Sessions with no identifiable $P(t)$ peaks and missed sessions are left blank.}
\label{atab:phaseshift2}
\centering
\begin{tabular}{llrrrrrrrrrrr}
\toprule
&          &  \multicolumn{6}{c|}{Continuous Paradigm} & \multicolumn{5}{|c}{Intermittent Paradigm} \\
Mvmt.  &  Subj.  &  Period  &  Elec.  &  Flex Lag  &  Extd Lag  &  Num Flex  &  Num Extd  &  Elec.  &  Flex Lag  &  Extd Lag  &  Num Flex  &  Num Extd  \\
\midrule
  PG  &  \sj{SJ0084}  &  1642 ms  &     G22  &    11$\pm$8 \%  &    4$\pm$10 \%  &    89 / 97  &   55 / 97  &     G20  &    -9$\pm$115 ms  &   -42$\pm$173 ms  &  23 / 23  &  22 / 22  \\
      &  \sj{SJ0087}  &  1155 ms  &  GRID64  &    11$\pm$6 \%  &     5$\pm$7 \%  &   96 / 102  &  18 / 102  &  GRID56  &     54$\pm$43 ms  &     30$\pm$98 ms  &  30 / 30  &  29 / 29  \\
      &  \sj{SJ0088}  &   975 ms  &   LPS12  &   26$\pm$11 \%  &   -4$\pm$11 \%  &   93 / 102  &   4 / 101  &   LFG04  &    155$\pm$93 ms  &    -86$\pm$53 ms  &  21 / 22  &  18 / 22  \\
      &  \sj{SJ0085}  &  1165 ms  &      F4  &    24$\pm$8 \%  &   21$\pm$13 \%  &    92 / 96  &   2 / 100  &      F4  &   169$\pm$322 ms  &  -256$\pm$439 ms  &  24 / 25  &  17 / 28  \\
      &  \sj{SJ0096}  &  1313 ms  &  Grid60  &    13$\pm$7 \%  &    -1$\pm$9 \%  &    29 / 30  &   10 / 28  &  Grid52  &    156$\pm$57 ms  &    -92$\pm$80 ms  &  25 / 25  &  25 / 25  \\
      &  \sj{SJ0082}  &   835 ms  &     PS5  &   10$\pm$11 \%  &                 &    86 / 99  &    0 / 94  & \multicolumn{5}{l}{Subject did not perform any intermittent movements.}  \\
\midrule
   W  &  \sj{SJ0084}  & \multicolumn{6}{l}{Subject did not perform the movement correctly.}                 &     G20  &  -162$\pm$244 ms  &  -206$\pm$216 ms  &  21 / 21  &  20 / 20  \\
      &  \sj{SJ0087}  &  1514 ms  &  GRID56  &     7$\pm$6 \%  &    -1$\pm$8 \%  &  111 / 118  &  45 / 119  &     GR6  &     75$\pm$65 ms  &     73$\pm$53 ms  &  25 / 27  &  25 / 25  \\
      &  \sj{SJ0088}  &  1894 ms  &   LFG04  &    3$\pm$11 \%  &     6$\pm$7 \%  &    54 / 68  &   60 / 66  &   LFG15  &  -201$\pm$114 ms  &  -240$\pm$195 ms  &  26 / 26  &  24 / 25  \\
      &  \sj{SJ0085}  &  1863 ms  &     F10  &    8$\pm$14 \%  &    1$\pm$17 \%  &    80 / 99  &   65 / 99  &     F10  &  -508$\pm$160 ms  &  -657$\pm$269 ms  &  18 / 21  &  18 / 22  \\
      &  \sj{SJ0096}  & \multicolumn{6}{l}{Subject did not perform the movement correctly.}                 &  Grid59  &    -33$\pm$84 ms  &   -53$\pm$205 ms  &  20 / 22  &  21 / 21  \\
\midrule
  PS  &  \sj{SJ0084}  &  2521 ms  &     G28  &     3$\pm$8 \%  &     2$\pm$7 \%  &    63 / 91  &   65 / 92  &     G28  &  -165$\pm$267 ms  &  -251$\pm$226 ms  &  22 / 23  &  23 / 23  \\
      &  \sj{SJ0087}  &  1500 ms  &     GR6  &    11$\pm$8 \%  &    11$\pm$7 \%  &   95 / 101  &  85 / 100  &     GR6  &     72$\pm$71 ms  &     74$\pm$68 ms  &  28 / 28  &  29 / 29  \\
      &  \sj{SJ0088}  &  1607 ms  &   LFG15  &    -3$\pm$9 \%  &    -6$\pm$8 \%  &    85 / 96  &   71 / 95  &   LFG15  &  -206$\pm$182 ms  &  -274$\pm$180 ms  &  28 / 28  &  29 / 29  \\
      &  \sj{SJ0096}  &  1609 ms  &  Grid59  &    -1$\pm$9 \%  &     4$\pm$7 \%  &    13 / 13  &   11 / 13  &  Grid59  &    -1$\pm$106 ms  &    -55$\pm$97 ms  &  22 / 23  &  22 / 23  \\
      &  \sj{SJ0082}  &  1847 ms  &     PS5  &     5$\pm$8 \%  &     6$\pm$5 \%  &      2 / 5  &     3 / 5  & \multicolumn{5}{l}{Subject did not perform any intermittent movements.}  \\
\midrule
   E  &  \sj{SJ0084}  &  2759 ms  &     G28  &  -30$\pm$14 \%  &   -26$\pm$7 \%  &    86 / 96  &   82 / 96  &     G28  &  -320$\pm$376 ms  &  -581$\pm$287 ms  &  24 / 25  &  25 / 25  \\
      &  \sj{SJ0087}  &  1468 ms  &     GR6  &   -35$\pm$8 \%  &  -33$\pm$13 \%  &   84 / 100  &  83 / 101  &     GR6  &   -98$\pm$197 ms  &   -32$\pm$246 ms  &  25 / 25  &  25 / 25  \\
      &  \sj{SJ0088}  &  \multicolumn{6}{l}{No identifiable $P(t)$ power bursts.}                           &   LFG15  &  -495$\pm$133 ms  &  -454$\pm$223 ms  &  26 / 26  &  25 / 26  \\
      &  \sj{SJ0096}  &  1675 ms  &  Grid59  &  -29$\pm$12 \%  &  -26$\pm$24 \%  &   92 / 100  &   78 / 97  &  Grid59  &  -241$\pm$240 ms  &  -112$\pm$253 ms  &  21 / 23  &  21 / 23  \\
      &  \sj{SJ0082}  &  1392 ms  &     PS5  &   -26$\pm$8 \%  &   -33$\pm$7 \%  &   100 / 99  &   78 / 99  &  \multicolumn{5}{l}{Subject did not perform any intermittent movements.} \\
\midrule
  SR  &  \sj{SJ0084}  &  2668 ms  &     G20  &  -29$\pm$12 \%  &  -22$\pm$13 \%  &    83 / 97  &   91 / 96  &     G20  &  -164$\pm$662 ms  &  -639$\pm$187 ms  &  21 / 22  &  20 / 21  \\
      &  \sj{SJ0087}  &  \multicolumn{6}{l}{No identifiable $P(t)$ power bursts.}                           &  GRID64  &   -90$\pm$390 ms  &  -388$\pm$344 ms  &  25 / 25  &  20 / 23  \\
      &  \sj{SJ0088}  &  2072 ms  &   LFG15  &  -13$\pm$14 \%  &    -8$\pm$9 \%  &   90 / 103  &  97 / 101  &   LFG15  &  -546$\pm$168 ms  &  -522$\pm$189 ms  &  25 / 25  &  26 / 27  \\
      &  \sj{SJ0096}  &  1991 ms  &  Grid59  &  -20$\pm$20 \%  &  -18$\pm$19 \%  &   85 / 101  &   89 / 98  &  Grid59  &  -384$\pm$334 ms  &  -288$\pm$291 ms  &  18 / 24  &  20 / 24  \\
\midrule
 SFE  &  \sj{SJ0084}  &  2442 ms  &     G28  &   -26$\pm$9 \%  &  -30$\pm$20 \%  &   98 / 101  &  96 / 101  &     G28  &  -605$\pm$195 ms  &   210$\pm$221 ms  &  10 / 24  &  20 / 24  \\
      &  \sj{SJ0087}  &  2201 ms  &     GR6  &  -20$\pm$11 \%  &  -29$\pm$17 \%  &  100 / 101  &  96 / 101  &     GR6  &  -245$\pm$153 ms  &   -22$\pm$242 ms  &  25 / 25  &  24 / 25  \\
      &  \sj{SJ0088}  &  2312 ms  &   LFG15  &  -16$\pm$13 \%  &  -21$\pm$19 \%  &    75 / 99  &   89 / 99  &   LFG15  &  -547$\pm$102 ms  &  -510$\pm$190 ms  &  21 / 26  &  26 / 26  \\
      &  \sj{SJ0096}  &  1933 ms  &  Grid59  &   -23$\pm$8 \%  &  -27$\pm$12 \%  &    95 / 99  &   78 / 97  &  Grid59  &  -521$\pm$260 ms  &   136$\pm$436 ms  &  20 / 28  &  22 / 28  \\
\bottomrule
\end{tabular}
\end{sidewaystable*}

\end{document}